\documentclass[a4paper,11pt]{article}
\usepackage{jheppub}

\usepackage{braket}
\usepackage{slashed}
\usepackage{booktabs}
\usepackage{mathrsfs}
\usepackage[mathscr]{euscript}

\def\Tr{\textrm{Tr}}
\def\hc{\textrm{h.c.}}

\title{\texorpdfstring
{\boldmath{Soft-Collinear Chiral Perturbation Theory \\
for $B \to \pi$ form factors at large recoil}}
{Soft-Collinear Chiral Perturbation Theory \\
for B to pi form factors at large recoil}
}

\abstract{
We develop an effective hadronic theory for heavy-meson decays into energetic pions, organized by the approximate heavy-quark and chiral symmetries. The dynamical degrees of freedom of the theory consist of quasi-static heavy-meson fields coupled to soft and collinear pions, relevant for exclusive $B \to \pi$ transitions at large recoil energy ${E_\pi \simeq M_B/2}$. Our approach exploits the factorization of soft and collinear quarks in Soft-Collinear Effective Theory (SCET), leading to an emergent duplication of chiral symmetry in the two kinematic sectors. Promoting the soft and collinear chiral transformations to local symmetries enables us to systematically match non-local four-quark operators in SCET --- governing the soft-overlap contribution to heavy-to-light form factors --- onto effective operators constructed from hadronic fields below the scale of chiral symmetry breaking. As a first application, we compute the logarithmically enhanced pion-mass dependence of the $B \to \pi$ form factors at large recoil by calculating the one-loop matrix elements of the leading hadronic operators. We find that the non-local transition currents in our effective theory lead to a pattern of chiral-logarithmic corrections that is not compatible with the existing formulation provided by Hard-Pion Chiral Perturbation Theory.

}

\preprint{P3H-26-075, 
SI-HEP-2026-23}

\author[a]{Thorsten Feldmann,}
\author[a]{Jack Jenkins,}
\author[a]{Jaime del Palacio Lirola}

\affiliation[a]{Theoretische Physik 1, Center for Particle Physics Siegen, and Cluster of Excellence “Color meets Flavor,”
Universität Siegen, 57068 Siegen, Germany}

\emailAdd{thorsten.feldmann@uni-siegen.de}
\emailAdd{jack.jenkins@uni-siegen.de}
\emailAdd{jaime.delpalaciolirola@uni-siegen.de}

\begin{document}
\maketitle
\flushbottom
\clearpage 


\section{Introduction}

The semileptonic decays $B\to\pi\ell\nu$ and $B_s\to K\ell\nu$ provide direct access to the magnitude of the CKM matrix element $|V_{ub}|$ and play a central role in precision tests of CKM unitarity within the Standard Model~\cite{Charles:2015gya,Bona:2006ah,HeavyFlavorAveragingGroupHFLAV:2024ctg}. Theoretical predictions for these processes are limited by the precision of hadronic heavy-to-light form factors, which can be computed from first principles in lattice QCD~\cite{FlavourLatticeAveragingGroupFLAG:2024oxs, Dalgic:2006dt, Colquhoun:2015mfa, FermilabLattice:2015mwy, Flynn:2015mha, Colquhoun:2022atw, Bouchard:2014ypa, FermilabLattice:2019ikx, Flynn:2023nhi}. The same kind of form factors, such as for $B \to K$, are also key ingredients in theoretical predictions for nonleptonic and rare semileptonic decays in factorization approaches, see e.g.\ Refs.~\cite{Beneke:1999br,Beneke:2000ry,Ali:1999mm,Beneke:2001at}.

At present, lattice calculations are restricted to the low hadronic recoil region, corresponding to a large invariant mass $q^2\simeq M_B^2$ of the leptons, and must be extrapolated to the large-recoil region to be used in the phenomenological analysis of the available experimental data~\cite{Belle:2010hep,Belle:2013hlo,BaBar:2010efp,BaBar:2012thb,LHCb:2020ist}. At very low pion energies the $B\to \pi$ form factors can also be described in the framework of heavy-meson chiral perturbation theory (HMChPT)~\cite{Wise:1992hn,Yan:1992gz,Burdman:1992gh,Boyd:1994pa,Casalbuoni:1996pg}. In particular, HMChPT allows one to calculate the non-analytic dependence on the pion mass from chiral loops. On the other hand, for large pion energies $E_\pi \simeq M_B/2$, corresponding to $q^2\ll M_B^2$, the constituents of the final hadronic state are highly relativistic, and the form factors can be estimated from light-cone sum rules, which are based on dispersion relations and quark-hadron duality in continuum QCD, see e.g.\ the reviews in Refs.~\cite{Colangelo:2000dp,Khodjamirian:2023wol}. In this approach, however, the necessary modeling of hadronic spectral densities induces an irreducible source of systematic uncertainty which limits the accuracy of theoretical predictions in this framework.

{An alternative approach to access heavy-to-light form factors at large recoil is based on the hard-scattering picture which developed in the context of QCD factorization~\cite{Beneke:1999br,Beneke:2000ry,Beneke:2000wa} and has been formalized in Soft-Collinear Effective Theory (SCET)~\cite{Bauer:2000yr,Bauer:2001yt,Beneke:2002ph,Beneke:2003pa,Lange:2003pk,Hill:2002vw}. The approach is based on the separation of momentum modes related to the scale hierarchies 
\begin{align}
\Lambda_{\rm QCD} \ll \sqrt{\Lambda_{\rm QCD} m_b}\ll m_b  \, . \label{eq:SCET-hierarchy}
\end{align}
The form factors at large hadronic recoil can be written as a sum of two terms~\cite{Beneke:2000wa},
\begin{align}
    f_i^{B \to \pi}(E_\pi)
    &\simeq
    C_i(E_\pi,\mu)\,\zeta_\pi(E_\pi,\mu)
    \nonumber \\
    &\hspace{1cm} + \int\limits_0^\infty d\omega \int\limits_0^1 du\;
    F_B(\mu) \, \phi_B^+(\omega,\mu)\,
    T_i(E_\pi,\omega,u,\mu)\,
    F_\pi \,\phi_\pi(u,\mu) \, .
    \label{eq:ffs_sum}
\end{align}
The second term, usually referred to as the ``hard-scattering contribution,'' factorizes into the leading two-particle light-cone distribution amplitudes (LCDAs) for the $B$-meson (denoted by $\phi_B^+$) and pion (denoted $\phi_\pi$), together with their respective decay constants $F_B$ and $F_\pi$, and a perturbatively calculable hard-scattering kernel $T_i$. In contrast, the so-called ``soft-overlap contribution'' in the first term does not admit a complete separation of the soft and collinear dynamics in the initial and final-state hadron, leaving a single form factor $\zeta_\pi$ as a universal hadronic input function, together with hard matching coefficients $C_i$. All ingredients in the factorization formula depend on an arbitrary factorization scale $\mu$, and are governed by the renormalization-group equations of non-local operators in SCET.

The aim of this paper is to invoke soft-collinear factorization in order to generalize HMChPT to the case of large recoil energies, enabling a consistent EFT description of the full set of
$B_{(s)} \to (\pi, K, \eta)$ form factors 
in this kinematic regime. To develop the formalism in the most transparent way, we focus on 
the $B \to \pi$ scalar and vector form factors. The relevant dynamics are governed by a combined heavy-quark/large-energy and chiral expansion, based on
\begin{align}
    m_\pi \ll 4\pi F_\pi \ll E_\pi\sim m_b \, ,
\end{align}
where $F_\pi \simeq 92.4 \ \text{MeV}$ is the pion decay constant. The expansion of the form factors around the chiral limit is generically expected to exhibit a logarithmic dependence on the light-quark masses of the form $\sim m_\pi^2 \ln m_\pi^2$. Concerning the LCDAs appearing in the \emph{factorizable} contribution to the form factors in Eq.~\eqref{eq:ffs_sum}, it is known that these so-called ``chiral logarithms'' enter as an overall multiplicative factor, leaving the dependence on the large energy scale in the hard-scattering kernels unaffected~\cite{Chen:2003fp}. Our goal in this paper is to determine the light-quark mass dependence of the \emph{non-factorizable} soft-overlap form factor, where -- as we will show -- the situation is somewhat more involved.

Another approach to the large-recoil region, known as Hard-Pion ChPT (HPChPT), was originally developed to compute chiral logarithms in amplitudes with energetic external pions~\cite{Flynn:2008tg,Bijnens:2009yr}, and was later applied to heavy-to-light form factors in Refs.~\cite{Bijnens:2010ws,Bijnens:2010jg} (see also Refs.~\cite{Colangelo:2012ew,Grinstein:2005ry} for similar work)}. The essential conclusion in Refs.~\cite{Bijnens:2010ws,Bijnens:2010jg} is that the leading chiral logarithms factorize from the energy dependence of the form factors, a result which has been employed in some lattice QCD analyses~\cite{FermilabLattice:2015mwy, Flynn:2015mha}. In this work we re-frame the problem by systematically matching QCD to a hadronic effective theory through a sequence of partonic EFTs, in particular taking into account the non-local nature of the relevant operators in SCET. The problem requires extending HMChPT by introducing a collinear hadronic sector for the energetic pions. Our approach is to exploit the separation of soft and collinear degrees of freedom in the SCET Lagrangian to identify an emergent duplication of chiral symmetry, which we promote to a local redundancy, in order to construct covariant currents involving hadronic fields. The dressing of heavy-light currents by soft and collinear chiral structures plays a role analogous to Wilson lines in gauge theory, reflecting the emergence of separate soft and collinear `hidden symmetries' of nonlinear realizations~\cite{Callan:1969sn}, without going so far as to introduce vector mesons as dynamical gauge fields~\cite{Bando:1987br}.

The structure of the paper is as follows. In Section~\ref{sec:2} we review the effective hadronic representation relevant for $B\to\pi\ell\nu$ at low recoil, emphasizing the realization of heavy-quark and chiral symmetries in terms of composite fields transforming in nonlinear representations of the chiral group. This provides the starting point for a systematic extension to the large-recoil regime. In Section~\ref{sec:3} we develop this extension by matching the relevant four-quark operators in SCET-2 below the hard-collinear scale onto operators constructed from heavy-meson fields and soft and collinear pions. In Section~\ref{sec:4} we apply the effective theory to compute the leading chiral logarithms of the soft-overlap form factor and analyze their interplay with the energy dependence. We conclude in Section~\ref{sec:5} with a discussion of implications for chiral extrapolations of heavy-to-light form factors on the lattice and possible extensions of the framework.


\section{Hadronic effective Lagrangians} 
\label{sec:2}


\subsection{Soft pions}

In this section we review the covariant formulation of the chiral Lagrangian, see e.g.\ Refs.~\cite{Coleman:1969sm, Callan:1969sn, Honerkamp:1971sh,Meetz:1969as}, as developed for light mesons and baryons, see e.g.\ Ref.~\cite{Gasser:1987rb}, and subsequently for heavy hadrons, see e.g.\ Refs.~\cite{Burdman:1992gh,Yan:1992gz,Wise:1992hn}. In quantum chromodynamics (QCD) the strong interactions of $N$ flavours of massless quarks $q = (q_1, \dots, q_N)^T$ coupled to external scalar, pseudoscalar, vector and axial-vector sources are described by a Lagrangian $\mathcal{L} = \mathcal{L}_\textrm{YM} + \mathcal{L}_q$, where $\mathcal{L}_\textrm{YM}$ is the Yang-Mills term for the gluons, and
\begin{align}
    \mathcal{L}_q &= \bar{q}(i\slashed{D} - s + ip\gamma_5 +\slashed{v} + \slashed{a}\gamma_5)q \nonumber \\
    & = \bar{q}_L(i\slashed{D} + \slashed{\ell})q_L + \bar{q}_R (i\slashed{D} + \slashed{r}) q_R  - \bar{q}_L \chi q_R - \bar{q}_R \chi^\dagger q_L\, . \label{eq:LQCD}
\end{align}
Here $iD_\mu = i\partial_\mu + g A_\mu^a T^a$ is the QCD covariant derivative, the chiral quark field projections are $q_L = P_L q$ and $q_R = P_R q$ with $P_{L(R)}=(1\mp \gamma_5)/2$, and the sources appear in the linear combinations $\chi^{(\dagger)} = s\mp ip$, $\ell^\mu = v^\mu-a^\mu$ and $r^\mu = v^\mu+a^\mu$. The Lagrangian is invariant under 
\emph{local} transformations in the chiral group $G \simeq SU(N)_L \times SU(N)_R$ specified by
\begin{align}
    q_L \mapsto L(x) \, q_L \, , \indent
    q_R \mapsto R(x) \, q_R \, , \label{eq:qTrans}
\end{align}
provided the external source fields transform as
\begin{align}
    \chi \mapsto L \chi R^\dagger \, , \indent
    \ell_\mu \mapsto L \ell_\mu L^\dagger + L i\partial_\mu L^\dagger \, , \indent
    r_\mu \mapsto Rr_\mu R^\dagger + R i\partial_\mu R^\dagger \, . \label{eq:extTrans}
\end{align}

In the context of coupling to heavy fields in Section~\ref{sec:hhcpt}, it will be convenient to perform a change of variables, introducing
\begin{align}
q_L' = \xi^\dagger q_L , \qquad q_R' = \xi \, q_R  \, .
\label{eq:qredef}
\end{align}
The matrices can be parameterized as
\begin{align}
    \xi(x)=\exp \left[\frac{i\pi^a(x) \, t^a}{F} \right] \, ,
\end{align}  
with the generators $t^a$ of the $su(N)$ algebra.\footnote{The generators are normalized by $\Tr[t^a t^b] = \delta^{ab}/2$.} Under chiral symmetry one then imposes the following transformation law~\cite{Coleman:1969sm},
\begin{align} 
\xi \mapsto L \xi U^\dagger \equiv U \xi R^\dagger\, , 
    \label{eq:utrans}
\end{align}
where the transformation matrix $U(x)=U(L,R,\xi)$
is implicitly defined by the equation on the r.h.s.~above. In consequence the primed quark fields transform as
\begin{align}
    \, q'_L \mapsto U q_L' \,, \quad  q'_R\mapsto Uq_R' \, .
\end{align}
For the diagonal sub-group of $SU(N)_L \times SU(N)_R$, one simply has $U=L=R$, such that the matrix field $\xi$ as well as the fields $\pi(x) = \pi^a(x) t^a$ transform homogeneously under $SU(N)_{L=R}$. The fields $\pi(x)$ can thus be identified with the $(N^2-1)$ Goldstone bosons, once the chiral $SU(N)\times SU(N)$ symmetry is spontaneously broken by a non-trivial vacuum.\footnote{Notice that the condensate in the new coordinates is invariant under the full symmetry group $G$.} 

After the field redefinitions, all explicit references to the matrices $L,R$ have been absorbed into the definition of the compensator $U$. The fermion Lagrangian coupled to Goldstone bosons
in the new coordinates is given by
\begin{align}
    \mathcal{L}'_q &= \bar{q}' (i\slashed{D} - S + iP\gamma_5 + \slashed{V} - \slashed{A}\gamma_5)q'\label{eq:Leff} \, . 
\end{align}
The auxiliary fields appearing in the scalar, pseudoscalar, vector and axial-vector currents are defined by
\begin{align}
    S &= \frac{1}{2}(\xi^\dagger \chi \xi^\dagger + \hc) \, , \quad P=\frac{i}{2}(\xi^\dagger \chi \xi^\dagger - \hc) \, ,  \nonumber \\
    V_\mu &= \frac{1}{2}(\xi^\dagger iD_\mu \xi + \xi i D_\mu \xi^\dagger) \, , \quad A_\mu = \frac{1}{2}( \xi^\dagger i D_\mu \xi - \xi i D_\mu \xi^\dagger) \, , \label{eq:chiraldef}
\end{align}
where $iD_\mu \xi = (i\partial_\mu+\ell_\mu) \xi$ and $i D_\mu \xi^\dagger = (i \partial_\mu + r_\mu)\xi^\dagger$
and transform as
\begin{align}
    S & \mapsto U S U^\dagger\, , \quad P \mapsto U P U^\dagger\, , \quad V_\mu \mapsto UV_\mu U^\dagger + i U \partial_\mu U^\dagger \, , \quad A_\mu \mapsto U A_\mu U^\dagger \, .
\label{eq:chiral_trans}
\end{align}

Finally, the effective action of the fields $\xi,\xi^\dagger$ coupled to the external sources is obtained by performing the path integral over the (primed) quark and gauge fields. The fermion path integral induced by Eq.~\eqref{eq:Leff} is Gaussian, and can be evaluated analytically in the form of a functional determinant. The remaining integral over the gauge fields (including the Yang-Mills term) is not Gaussian and cannot be performed perturbatively at the scales relevant to the effective action. However, the effective action can only depend on the Hermitian building blocks $S(\xi,\xi^\dagger), P(\xi,\xi^\dagger), V^\mu(\xi, \xi^\dagger), A^\mu(\xi, \xi^\dagger)$ appearing in the functional determinant in Eq.~\eqref{eq:Leff}. Restricting ourselves to low-energy scattering of Goldstone bosons, the effective Lagrangian can be constructed as a tower of composite operators which are sorted by {\it chiral power counting}, where the leading term can be written as
\begin{align}
  \mathcal{L}_\chi^{(2)} &= F^2\big( \textrm{Tr}[A_\mu A^\mu] + B_0 \Tr[S] \big) \, , \label{eq:Lchi2}
\end{align}
where $F$ and $B_0$ are constants.\footnote{Note that an equivalent way of writing the chiral effective
Lagrangian is in terms of the field $\Sigma=\xi^2$, with $\Tr[A_\mu A^\mu] = \frac{1}{4} \Tr[D_\mu \Sigma^\dagger D^\mu \Sigma]$.}
The constant $F$ can be determined in terms of the pion decay constant by calculating the Noether currents associated to the broken symmetry generators, which results in $F \simeq F_\pi =f_\pi/\sqrt{2}$ at lowest order. Moreover, if one includes the explicit breaking of chiral symmetry by a (small) quark-mass matrix $m_q$, this can be accounted for by the scalar source term which results in a finite (pseudo-)Goldstone-boson mass, $m_\pi^2 = 2B_0 m_q$.


\subsection{Heavy mesons}
\label{sec:hhcpt}


We are now in a position to couple the (pseudo-)Goldstone bosons to heavy-meson
fields. In the heavy-quark limit, the lightest heavy-meson states form 
a spin multiplet, consisting of a pseudoscalar component $P_v(x)$ (to be identified with e.g.\ a $\bar B$-meson) and a vector component $P_v^*(x)$  (to be identified with e.g.\ a $\bar B^*$ meson), which we denote as
\begin{align}
H_v = \frac{1+ \slashed{v}}{2}\big(\slashed{P}_v^{*} + iP_v\gamma_5\big), \quad \bar H_v = \gamma^0H^\dagger_v \gamma^0 .
\end{align}
Using the same field-dependent compensator $U(L,R,\xi)$ that appears in the transformation of $\xi$ in Eq.~\eqref{eq:utrans}, the heavy field is taken to 
transform under $SU(N)_L\times SU(N)_R$ as ${H_v \mapsto H_v U^\dagger}$, and under the unbroken vector subgroup one simply has $U=V$ and ${H_v \to H_v V^\dagger}$. The leading terms of the strong interaction Lagrangian are then
\begin{align}
\mathcal{L}_H^{(1)}
= -\Tr \big[\bar{H}_v i v\cdot D H_v \big] + g_\pi \Tr \big[\bar{H}_{v} H_v \gamma_\mu \gamma_5 A^\mu \big]\,,
\label{eq:HMChPT_L}
\end{align}
where $D^\mu H_v = \partial^\mu H_v + i H_vV^\mu$, and the derivative acting on the $H_v$ field picks up only the residual momentum of the heavy meson. In this case, the traces should be understood to be over both Dirac and flavor indices.

For the matching of an external left-handed heavy to light current, the hadronic operator must transform under $SU(N)_L\times SU(N)_R$ as $(\bar{N}_L,1_R)$ and as a doublet under heavy-quark spin symmetry. At leading power, it is uniquely given by
\begin{align}
\bar q_L\Gamma h_v = \frac{a}{2}\,\Tr\big[H_v P_R \Gamma \big]\xi^\dagger .
\label{eq:HLcurrent}
\end{align}
Taking the $B$-meson-to-vacuum matrix element of both sides, the parameter $a$ is identified as the HQET decay constant $a = \sqrt{M_{P}}\,F_{P}$, with the appropriate normalization convention.


\subsection{Collinear pions}


The effective HMChPT Lagrangian constructed so far 
allows one to calculate decays of heavy $B$-mesons
into an arbitrary number of \emph{soft} Goldstone bosons, i.e.\ pions with small energy in the $B$-meson rest frame.
This includes corrections from {\it chiral loops} which determine the leading non-analytic dependence on the 
light-meson masses, see e.g.\  \cite{Fleischer:1992tn,Falk:1993fr,Bijnens:2010ws}
for the $B \to \pi$ form factors at low recoil.
In order to quantify the analogous effects 
in the high-recoil region, i.e.\ processes with \emph{energetic} pions, it is necessary to generalize the standard HMChPT
framework. To this end, we will combine concepts of ChPT and Soft-Collinear Effective Theory (SCET).

We now start with the leading SCET-2 strong-interaction Lagrangian in the presence of external sources, which can be written as $\mathcal{L} = \mathcal{L}_{\textrm{YM}}+\mathcal{L}_c + \mathcal{L}_s\, . $
The Yang-Mills Lagrangian involves only the (soft and collinear) gluon fields, together with the gauge-fixing terms, which are not critical to the following analysis since they do not transform under chiral symmetry. The soft and collinear quark terms have the generic form as in Eq.~\eqref{eq:LQCD}:
\begin{align}
\mathcal{L}_s &= \bar{q}_{s,L}(i\slashed{D} + \slashed{\ell}_s)q_{s,L} + \bar{q}_{s,R} (i\slashed{D} + \slashed{r}_s) q_{s,R} - \bar{q}_{s,L} \chi_s q_{s,R} - \bar{q}_{s,R} \chi^\dagger_s q_{s,L} \, ,\\
    \mathcal{L}_c
    &= \bar{q}_{c,L}(i\slashed{D} + \slashed{\ell}_c)q_{c,L} + \bar{q}_{c,R} (i\slashed{D} + \slashed{r}_c) q_{c,R} - \bar{q}_{c,L} \chi_c q_{c,R} - \bar{q}_{c,R} \chi^\dagger_c q_{c,L} \, ,
\end{align}
in terms of the QCD covariant derivatives. The chiral group $(SU(N)_L \times SU(N)_R)$ of QCD is therefore enlarged to $$(SU(N)_L \times SU(N)_R)_s \times(SU(N)_L \times SU(N)_R)_c$$ in SCET-2, where
\begin{align}
    q_{c,L} \mapsto L_c q_{c,L} \, , \quad q_{c,R} \mapsto R_c q_{c,R} \, , \quad
    q_{s,L} \mapsto L_s q_{s,L} \, , \quad q_{s,R} \mapsto R_s q_{s,R} \, ,
\end{align}
and $L_s \neq L_c$ and $R_s \neq R_c$ are distinct transformations on the soft and collinear quarks. Accordingly, the generalization of Eq.~\eqref{eq:extTrans} results in
\begin{align}
    \ell_s^\mu \mapsto L_s \ell_s^\mu L_s^\dagger + L_s i \partial^\mu L_s^\dagger \, , \quad \ell_c^\mu \mapsto L_c \ell_c^\mu L_c^\dagger + L_c i \partial^\mu L_c^\dagger \, ,\label{eq:extTransSCET}
\end{align}
and analogously for the other external currents.

We should recall that the local symmetry in Eq.~\eqref{eq:extTransSCET}  is equally as artificial as in the QCD case Eq.~\eqref{eq:extTrans}. In either case, the external fields are fixed, potentially explicitly breaking chiral symmetry, and their transformations are mainly useful to restore an exact chiral symmetry in the EFT, which may reduce the number of independent operators at a given order in the power counting (which will be introduced for the hadronic theory shortly). 

Having specified the transformation of the external fields, the same procedure that was 
applied to QCD in Eq.~\eqref{eq:qredef}, in which the pions are introduced via unitary field redefinitions of the quark fields, can be repeated for both soft and collinear sectors separately:
\begin{align}
    q_{c,L}' = \xi_c^\dagger q_{c,L} \, , \quad q_{c,R}' = \xi_c q_{c,R} \, , \quad q_{s,L}' = \xi_s^\dagger q_{s,L} \, , \quad q_{s,R}' = \xi_s q_{s,R} \, .
\end{align}
By rotating 
soft and collinear quarks separately, we have introduced both soft and collinear pions into the dynamics, which was our original motivation. By extension, one obtains soft and collinear variants of the derivative representations $V^\mu_a,A^\mu_a$ and sources $S_a, P_a$ as defined in Eq.~\eqref{eq:chiraldef}, now labeled by $a=s,c$, with which to construct the hadronic Lagrangian.

After (formally) integrating out the quarks and gluons, we are left with a hadronic theory in which the strong interactions within the collinear sector are described by the chiral Lagrangian with the same power-counting rules as for standard ChPT, whereas the soft sector (including the soft heavy meson fields) is described by the standard HMChPT. In summary, the effective Lagrangian has the form\footnote{From the bottom up perspective, the duplicated soft and collinear sectors would in principle allow independent leading-order constants. However, these constants are related to universal hadronic parameters in QCD which are defined in a boost invariant manner.}
\begin{align}
\mathcal{L}&=-\Tr \big[\bar{H}_v i v\cdot D H_v \big] + g_\pi \Tr \big[\bar{H}_{v} H_v \gamma_\mu \gamma_5 A^\mu_s \big] + F^2 \left( \Tr \big[A_{s\mu} A_s^\mu \big] +B_0 \Tr\big[S_s \big]\right) \nonumber \\
&\quad +F^2 \left(\Tr\big[A_{c\mu}A_c^\mu\big] +B_0 \Tr\big[S_c\big]\right) \, . \label{eq:LSCHMChPT}
\end{align}
By construction, the hadronic Lagrangian consists of two decoupled soft and collinear sectors. Both types of fields describe particles with low virtualities of order $m_\pi^2$, which corresponds to a SCET-2 setup in the hadronic theory. 

Finally, we note that interactions between soft and collinear pions are mediated by off-shell modes that are integrated into the matching coefficients of subleading operators in the hadronic theory. These descend from subleading Lagrangians in SCET-2, which are beyond the scope of this analysis.


\section{Matching of external currents}
\label{sec:3}


\subsection{Chiral symmetry of four-quark operators in SCET-2}

The hadronic representation of heavy-to-light currents follows after the two step matching $\textrm{QCD} \to \textrm{SCET-1} \to \textrm{SCET-2}$, which decouples the hard and hard-collinear scales. In the following we outline this partonic matching procedure, emphasizing the chiral structure of the currents and four-quark operators in the effective theories. In this context, one introduces the usual light cone decomposition of a momentum vector
\begin{align}
    k_\mu = \bar{n}\cdot k \frac{n_\mu}{2} + n\cdot k \frac{\bar{n}_\mu}{2} + k^\perp_\mu \, ,
\end{align}
where $n^2=\bar{n}^2=0$, $n\cdot \bar{n}=2$ and $n\cdot k^\perp = \bar{n}\cdot k^\perp =0$.

At the hard scale $\mu\sim m_b$, the QCD heavy-to-light current with a left-handed light quark is matched perturbatively onto SCET-1 as
\begin{align}
    &\bar{q}_{L,i} \Gamma b = \sum_A \int ds \,K_{\Gamma} ^{(A)}(s,E,\mu) \, J_i^{(A)}(s) + \sum_B \int ds \,dr \, K_{\Gamma}^{(B)}(s,r,E,\mu) \, J_i^{(B)}(s,r) \, . \label{eq:QCD_to_SCETI}
\end{align}
Following the notation of Ref.~\cite{Lange:2003pk}, the tensor coefficients $K_\Gamma$ absorb the Wilson coefficients which are perturbative and encode the dependence on the hard scales $m_b,E$. The first term in Eq.~\eqref{eq:QCD_to_SCETI} defines the set of two-body SCET-1 currents $J_i^{(A)}(s)=\bar{\xi}_{L,i}(s\bar{n}) \Gamma^{(A)} h_v(0)$, while the second term corresponds to leading-power three-body currents involving an explicit hard-collinear gluon~\cite{Bauer:2000yr,Beneke:2000wa, Bauer:2001yt,Beneke:2003pa,Lange:2003pk}. Both classes contribute at the same order in the $1/m_b$ expansion, but we focus on the type-A currents since our main interest is the soft-overlap contribution. 

\begin{figure}[t]
  \centering
  \includegraphics[height=4.5cm]{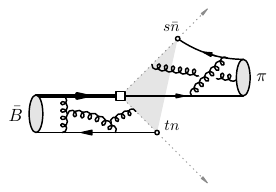}
  \hspace{1cm}
  \includegraphics[height=4.5cm]{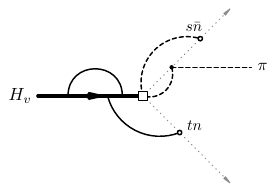}

  \caption{Illustration of the four-quark operators in SCET-2 exhibiting non-factorizable light-cone dynamics (left). The grey region is conventionally represented as a single vertex, with the light-cone non-locality of the associated SCET-2 operator left implicit. The corresponding chiral dynamics in the hadronic picture is illustrated on the right; soft and collinear pions are indicated by solid and dashed lines, respectively.}
  \label{fig:combined}
\end{figure}

The soft overlap form factor can be defined as the matrix element of the type-A scalar current
\begin{align}
    \bra{\pi^a(p)} \bar{\xi}_{L,i} h_v\ket{\bar B_j(v)} \equiv 2E_\pi \, t_{ji}^a \,\zeta_\pi(E_\pi,\mu) \, , \label{eq:overlapdef}
\end{align}
where $E_\pi=\bar{n}\cdot p/2$ and $t^a$ are the generators of chiral symmetry. The matrix element on the LHS of Eq.~\eqref{eq:overlapdef} is equivalent to the matrix element of the type-A term of the decomposition of Eq.~\eqref{eq:QCD_to_SCETI} in the heavy quark limit~\cite{Beneke:2003pa}. The conversion of the soft spectator into a collinear quark arises from time-ordered products of a type-A current with the subleading SCET-1 Lagrangian, as represented in Fig.~\ref{fig:combined}. After integrating out hard-collinear modes, the resulting objects are matched onto four-quark operators in SCET-2,
\begin{align}
i \int d^4 x \,T \big\{ J_i^{(A)}(0) 
\,\mathcal{L}^{(1)}_{mn}(x)\big\}
&=
\sum_{\gamma}\int ds\,dt\;
D^{(\gamma)}(s,t,E,\mu)\;\big[O_{imn}^{(\gamma)}(s,t)\big] \, ,
\label{eq:typeA_Tprod_to_SCET2}
\end{align}
where $\gamma$ labels a complete basis of Dirac and color structures, and $D^{(\gamma)}$ are a set of perturbative matching coefficients. We have applied translation operators to make the heavy-to-light current local. The flavor indices $m,n$ will be contracted when computing physical matrix elements, but for our purposes it is helpful to tensorize the operator, since we are about to manipulate these indices differently under soft and collinear chiral symmetries. To be precise, Eq.~\eqref{eq:typeA_Tprod_to_SCET2} should be understood with an implicit contraction (on both sides) with an external spurion field which takes a vacuum expectation value $\sim \delta_{mn}$ when the sources are set to zero. In the matrix element that enters the soft-overlap form factor
\begin{align}
    2E_\pi \, t_{ji}^a \,\zeta_{\pi}(E_\pi,\mu) = \sum_{\gamma}\int ds\,dt\;
D^{(\gamma)}(s,t,E_\pi,\mu) \,\braket{\pi^a(p)|O_{imn}^{(\gamma)}(s,t)|\bar B_j(v)}\, \delta_{mn}\,,
\label{eq:messengers}
\end{align} 
the soft and collinear indices of the spectator are appropriately contracted. 

The operators in Eq.~\eqref{eq:typeA_Tprod_to_SCET2} in an `un-Fierzed' basis have the form $\sim [\bar{\mathscr{X}}_{L,i}(0) \Gamma_M \mathscr{H}_v(0)] \times [\bar{\mathscr{Q}}_{j}(tn) \Gamma_N \mathscr{X}_{k}(s\bar{n})]$. In the massless limit, the chirality of the soft field ($\bar{ \mathscr{Q}}_{j}$) must be the same as the chirality of the collinear field ($\mathscr{X}_k)$, which is effectively equivalent to the condition $\{ \Gamma_N ,\gamma_5\}=0$ on the Dirac structure $\Gamma_N$. In consequence, in the Fierzed basis of Ref.~\cite{Lange:2003pk}, the color-singlet\footnote{Color-octet operators contribute to the anomalous dimension of the color-singlet ones, but do not have physical $B \to \pi$ matrix elements.} operators generated in the matching of the time ordered product of the scalar current and subleading Lagrangian to four-quark operators in SCET-2 take the form
\begin{align}
    O_{ijk}^{(1)} &= \,\bigl[\bar{\mathscr{X}}_{L,i}(0)\,\frac{\slashed{\bar n}}{2}\,\mathscr{X}_{L,k}(s\bar n)\bigr]\,
   \bigl[\bar{\mathscr{Q}}_{L,j}(tn)\,\frac{\slashed{\bar n}\slashed n}{4} \,\mathscr{H}_v(0)\bigr], \label{eq:O1}\\
    O_{ijk}^{(2)} &= \,\bigl[\bar{\mathscr{X}}_{L,i}(0)\,\frac{\slashed{\bar n}}{2} \, i\slashed{D}_{\perp} \mathscr{X}_{R,k}(s\bar n)\bigr]\,
   \bigl[\bar{\mathscr{Q}}_{R,j}(tn)\,\frac{\slashed n}{4} \mathscr{H}_v(0)\bigr], \\
   O_{ijk}^{(3)} &= \,\bigl[\bar{\mathscr{X}}_{L,i}(0)\,\frac{\slashed{\bar n}}{2} \, \slashed{\mathscr{A}}_{c\perp}(r\bar{n}) \mathscr{X}_{R,k}(s\bar n)\bigr]\,
   \bigl[\bar{\mathscr{Q}}_{R,j}(tn)\,\frac{\slashed n}{4} \mathscr{H}_v(0)\bigr], \\
   O_{ijk}^{(4)} &= \,\bigl[\bar{\mathscr{X}}_{L,i}(0)\,\frac{\slashed{\bar n}}{2} \mathscr{X}_{L,k}(s\bar n)\bigr]\,
   \bigl[\bar{\mathscr{Q}}_{L,j}(tn) \slashed{\mathscr{A}}_{s\perp}(un) \frac{\slashed n}{4} \mathscr{H}_v(0)\bigr] \, .\label{eq:O4}
\end{align}
The derivative in $O^{(2)}$ includes the spurion field $r^\mu_c$ inherited from the original QCD Lagrangian and does not include any gluon field. Under both the collinear and soft local chiral transformations,
\begin{align}
O_{ijk}^{(1,4)} &\mapsto L_{c,km}(s\bar n)\ O_{lnm}^{(1,4)} \ L_{c,li}^{\dagger}(0)\,
   L_{s,nj}^{\dagger}(tn) \, , \label{eq:O1trans} \\
    O_{ijk}^{(2,3)} &\mapsto R_{c,km}(s\bar n) \ O_{lnm}^{(2,3)} \ L_{c,li}^{\dagger}(0)\, R_{s,nj}^{\dagger}(tn) \, . \label{eq:O14trans} 
\end{align}
In the chiral limit, two classes of operators appear since the spectator quark can be left- or right-handed.

It is to be stressed that the renormalization of the operators in Eqs.~\eqref{eq:O1}-\eqref{eq:O4} requires a proper treatment of endpoint divergences. In Refs.~\cite{Becher:2003qh,Lange:2003pk}, for instance, this is formally achieved by taking into account operators with additional (unphysical) soft-collinear messenger modes. As the messenger modes will not change the chiral structure of the relevant SCET-2 operators, their effect can be absorbed into the matching coefficients of the hadronic theory. As a result, the non-factorization of the soft and collinear sectors in SCET-2 perturbation theory will be reflected in the non-factorizing dependence of the hadronic matching coefficients on the soft and collinear light-cone separations, as we will discuss further below.

\subsection{Hadronic building blocks and power counting}

To introduce a systematic power counting in the hadronic theory in the presence of an external current that introduces a large recoil energy $E_\pi \simeq M_B/2$, we need to account for the boost factor in soft-collinear interactions. Using the short-hand notation $k = (\bar{n}\cdot k,n\cdot k,k_\perp)$, the two relevant momentum modes scale like 
\begin{align}
    k_s \sim (p, p, p), \quad
    k_c\sim (E_\pi,p^2/E_\pi, p),
\end{align}
where $p$ is a soft-momentum in the $B$-meson rest frame. We can now define two separate power counting parameters $\epsilon = p/(4\pi F_\pi) \ll 1$ and $\lambda=p/E_\pi \ll 1$, and write
\begin{align}
    k_s \sim 4\pi F_\pi(\epsilon, \epsilon, \epsilon)
    = E_\pi \,  (\lambda ,\lambda ,\lambda ), \quad
    k_c \sim 4\pi F_\pi(\epsilon/\lambda, \epsilon\lambda, \epsilon)
    = E_\pi \, (1, \lambda^2,\lambda).
\end{align}
Here $\lambda$ plays the same role as the power counting parameter in SCET-2. We find it convenient to show both $4\pi F_\pi$ and $E_\pi\sim M_B/2$ as the reference scale, to show how the scaling of momenta is consistent with the usual power counting in both HMChPT and SCET. In particular, both terms in the expansion of the collinear kinetic term
\begin{align}
    A_c^2=\frac{1}{2} \{\bar{n}\cdot A_c,n\cdot A_c\} + A_{c\perp}^2 \, ,
\end{align}
have the same scaling $\sim p^2$. Therefore, no further expansion of the collinear pion Lagrangian in Eq.~\eqref{eq:LSCHMChPT} is required, and the propagator has the generic form of that of a complex, massive scalar field. Even at higher orders in the chiral Lagrangian, the potentially problematic component $\bar{n}\cdot A_c \sim E_\pi$ is always multiplied by $n \cdot A_c \sim p^2/E_\pi$ due to Lorentz invariance, restoring the usual chiral power counting.

Additionally, the nonlocal transformations in Eq.~\eqref{eq:O1trans} require a chiral connection to transport the local symmetry among composite operators in the hadronic theory. One could build two separate Wilson lines using the spurion fields $\ell^\mu$ and $r^\mu$ to transport the chiral symmetry between operators defined at different space-time points. However, the vector fields $V_\mu$ transform as gauge fields with respect to chiral symmetry as in Eq.~\eqref{eq:chiral_trans}, so it is sufficient to define one `chiral Wilson line' for each sector as
\begin{align}
    W_s(x) &\equiv \mathrm P \,\exp \left[ i \int_{-\infty}^{0} dt\,  n \cdot V_s(tn + x)\right] \mapsto U_s(x) W_s(x) U_s^\dagger(-\infty)\, , \\
    W_c(x) &\equiv \mathrm P \,\exp \left[ i \int_{-\infty}^{0} ds\, \bar n \cdot V_c(s\bar n + x)\right] \mapsto U_c(x) W_c(x) U_c^\dagger(-\infty) \, ,
\end{align}
that act as the necessary gauge links in our current.\footnote{Alternatively, if one defines two Wilson lines using the $V-A$ and $V+A$ chiral blocks, they encounter a reorganization of the theory that does not simplify the problem.} While $\bar{n}\cdot V_c$ and $n \cdot V_s$ can be written in terms of Wilson lines, the remaining projections of the vector fields, as well as the axial fields, appear in building blocks of the form
\begin{align}
\mathcal{V}^\mu_a&\equiv W_a^\dagger iD_a^\mu W_a \,, \qquad
\mathcal{A}_a^\mu \equiv W_a^\dagger A_a^\mu W_a \, ,
\end{align}
for soft and collinear pion fields separately $(a=s,c)$. Note that $\bar{n}\cdot \mathcal{V}_c = 0$ and $n \cdot \mathcal{V}_s=0$ following the defining equations of the Wilson lines $(i\bar{n}\cdot D)W_c=0$ and $(in\cdot D)W_s=0$. The remaining building blocks are
\begin{align}
    \Xi_{a,L} &\equiv \xi_a W_a\, , \, \quad
    \Xi_{a,R} \equiv W_a^\dagger\xi_a  \, , \quad
    \mathcal{H}_v \equiv  H_v W_s \, , \quad \mathcal{S}_a \equiv W_a^\dagger S_a W_a\,.
    \label{eq:dressed}
\end{align}
The building blocks (e.g.\ $\mathcal{A}$) transform as their corresponding primitives (e.g.\ $A$), but with the local transformations transported to the boundary: $U_a(x) \to U_a(-\infty)$. The boundary transformations enforce a subtle constraint on the contraction of the chiral indices, as $U_s(-\infty) U_c^\dagger(-\infty) \neq 1$. The condition for contracting Wilson lines belonging to the same sector (soft or collinear) is equivalent to writing the hadronic representation of the four-quark operator as a product of collinear and soft terms.

The basis above (including derivatives of those objects) is unique up to field redefinitions, and the scaling of all of the objects that can appear in our hadronic theory is summarized in Table~\ref{tab:powercounting}. Derivatives of collinear building blocks in the $\bar n$ direction are just a redundancy of light-cone separation, since the sum of such unsuppressed derivatives can always be expressed as light-cone transported fields
\begin{align}
  \int ds\,C(s)\,\phi_c(x+s\bar n) = \sum_{m=0}^{\infty}
  \frac{1}{m!}
  \left(
    \int ds\,s^m \, C(s)
  \right)
  (\bar n\cdot\partial)^m\phi_c(x) \, ,
\end{align}
where $\phi_c$ is a generic collinear field and $C(s)$ is a stand-in for the coefficients that will always accompany our operators in a convolution. Further derivatives of the non-local operators are also redundant, since by integration by parts they can be moved onto the coefficient, which anyways remains unknown. Therefore, we do not need to consider these as independent operators in our basis. However, derivatives of collinear building blocks in the $\perp$ and $n$ directions constitute independent basis elements.

\begin{table}[t]
\centering
\renewcommand{\arraystretch}{1.25}
\setlength{\tabcolsep}{10pt}

\begin{tabular}{@{}r@{\,}c@{\,}l @{\hspace{2em}} r@{\,}c@{\,}l@{}}
\toprule
\multicolumn{3}{c}{Soft} & \multicolumn{3}{c}{Collinear} \\
\midrule
 $\Xi_s$ & $\,\sim\,$ & $1$
  & $\Xi_c$ & $\,\sim\,$ & $1$ \\
$\mathcal{A}_s,\mathcal{V}_s$ & $\,\sim\,$ & $p$
  & $\bar n\cdot \mathcal{A}_c$ & $\,\sim\,$ & $E_\pi=p/\lambda$ \\
$\mathcal{H}_v$ & $\,\sim\,$ & $1$
  & $\mathcal{A}_c^\perp,\mathcal{V}_c^\perp$ & $\,\sim\,$ & $p$ \\
$D_\mu \mathcal{H}_v$ & $\,\sim\,$ & $p$
  & $n\cdot \mathcal{A}_c,\,n\cdot \mathcal{V}_c$ & $\,\sim\,$ & $p^2/E_\pi=p\lambda$ \\
$\mathcal{S}_s$ & $\,\sim\,$ & $p^2$
  & $\mathcal{S}_c$ & $\,\sim\,$ & $p^2$ \\
\bottomrule
\end{tabular}

\caption{Power counting of hadronic building blocks, with $\epsilon = p/(4\pi F_\pi)$ and $\lambda = p/E_\pi$. Derivatives acting on these building blocks scale as momenta in their corresponding sector, e.g.\ $(n \cdot \partial)(n\cdot \mathcal{A}_c) \sim (p\lambda)(p \lambda)$. In a given diagram, both soft and collinear chiral loops count as $\sim \epsilon^2$ due to sharing the same scalar propagator as ChPT.}
\label{tab:powercounting}
\end{table}


\subsection{Matching four-quark operators to the hadronic theory}
\label{sec:mainoperator}


The matching of the external currents from the partonic theory to the hadronic theory is done following the same logic as in standard HMChPT, shown in Eq.~\eqref{eq:HLcurrent}. However, the symmetries we have to comply with are more complex. Due to how the flavor indices are contracted, and the presence of non-localities, the operators in the hadronic theory must involve soft and collinear currents which transform under chiral symmetry at different space-time points.

The four-quark operator $O^{(1)}$ in Eq.~\eqref{eq:O1} will be our starting point for matching to the hadronic theory, as a more explicit example. It must be matched to the product of two currents that transform like
\begin{align}
    J_s^{(1)} &\mapsto J_s^{(1)}L_s^{\dagger}(tn)\, , \quad J_c^{(1)} \mapsto L_c(s\bar n)J_c^{(1)} L_c^\dagger(0) \, . \label{eq:Jctrans}
\end{align}
To first match the chiral structure associated to the collinear current $\bar{\mathscr{X}}_L \slashed{\bar n}\, \mathscr{X}_{L}$ to the hadronic theory, we consider hadronic operators consistent with the chiral symmetry in Eq.~\eqref{eq:Jctrans}. We begin with its simplest representative, namely the operator with a single insertion\footnote{The contribution to the $B \to \pi_c$ matrix element of the current  with zero insertions of the axial field $J_{c,0}^{(1)}$ can be absorbed into that of $J_{c,1}^{(1)}$, as shown in Appendix \ref{Appendix:J0_redundancy}.} of the axial field at an arbitrary point on the light cone,
\begin{align}
    J_{c,1}^{(1)}(s,s_1) = \Xi_{c,L}(s\bar n) \times \bar n\cdot \mathcal{A}_c (s_1\bar n)\times \Xi_{c,L}^\dagger(0)  \, ,\label{eq:J1def}
\end{align}
with an implicit integration over $s_1$, and the $\times$ indicates matrix multiplication. The effective insertion of the axial field at an arbitrary point on the light-cone is $ds_1 \times\bar n\cdot \mathcal{A}_c(s_1 \bar{n}) \sim 1$, i.e.~unsuppressed, since the scaling of $\bar{n} \cdot \mathcal{A}_c \sim E_\pi$ is compensated by an implicit integration over $ds_1\sim E_\pi^{-1}$. Therefore the collinear current defined in Eq.~\eqref{eq:J1def} may not be the unique leading current, since
$\bar{n}\cdot \mathcal{A}_c$ can be inserted in a candidate current any number of times like
\begin{align}
    J_{c,\alpha}^{(1)}(s,s_1,\dots, s_a) &= \Xi_{c,L}(s\bar n) \times \bar{n}\cdot \mathcal{A}_c(s_1\bar{n})\times \cdots \times \bar{n}\cdot \mathcal{A}_c(s_a\bar{n}) \times \Xi_{c,L}^\dagger(0) \, ,\label{eq:J2def}
\end{align}
with implicit integration over $s_1,\dots s_a$ with an unknown weight function. The index $\alpha$ in Eq.~\eqref{eq:J2def} labels all different currents, including singlet traces like $\Tr[\,\bar n\cdot\mathcal{A}_c(s_b\bar n)\times\cdots\times \bar{n}\cdot\mathcal{A}_c(s_c\bar n)\,]$ that can also be inserted to form valid operators (note $\Tr [\, \bar n \cdot \mathcal{A}_c(s\bar n)]=0$), while the index $a$ labels the number of extra non-localities.

For the soft sector current, the leading hadronic operator that transforms appropriately under chiral and heavy quark symmetries must have the form
\begin{align}
    \Tr [P_{R}\Gamma_{0}\mathcal{H}_v(0) \Gamma ] \, \Xi_{s,L}^{\dagger}(tn) \, ,
\end{align}
where $P_{R} \Gamma_0$ is the original structure in the SCET-2 partonic current \cite{Grinstein:2005ry}. All possible $\Gamma$ structures reduce to a unique trace
\begin{align}
    J_{s}^{(1)}(t) &= \mathrm{Tr}[\mathcal{H}_v(0)\slashed{n} (1+\gamma_5)] \, \Xi_{s,L}^\dagger(tn) \, . \label{eq:Jsdef}
\end{align}
Using $v \cdot P_v^*=0$ and $n \cdot v=\bar n \cdot v =1$, the trace reduces to
\begin{align}
    \Tr[\mathcal H_v(0)\slashed n (1+\gamma_5)] &= 2\bigl[n\cdot P_v^*(0)-iP_v(0)\bigr]W_s(0).
\end{align}
Eq.~\eqref{eq:Jsdef} is the non-local analogue of the current in HMChPT for $B \to \pi$ at low recoil in Eq.~\eqref{eq:HLcurrent}. Moreover, one should not expand in a series of local operators following $\Xi_{s,L}(tn)=\Xi_{s,L}(0)+\dots$, since each term in the expansion would not satisfy the local transformation of Eq.~\eqref{eq:Jctrans}.

Operators with additional powers of the soft fields $\mathcal{A}^\mu_{s} / \mathcal{V}^\mu_{s}$ will be suppressed, in a similar way as in the standard ChPT approach. In our case, the most general insertion of a soft field would be of the form
\begin{align}
    \int dt' \, C'(t,t',...) \, J_s'(t,t') \equiv \int dt' \, C'(t,t', ...) \, [...](0)[\mathcal{A}_s^\mu](t'n)[...](tn) \, .
\end{align}
The key fact is that the $t'$ dependence of the soft fields in the EFT is controlled by typical wave-lengths of order $\Delta t' \sim 1/p$, while the $t'$ dependence of the coefficient function $C'(t,t', ...)$ is controlled by the UV cut-off of the EFT, $\Delta t' \sim 1/(4\pi F_\pi)$. Following the standard reasoning for multi-pole expansion, the integration measure in the above convolution should correspond to the \emph{smallest} wave-length (i.e.\ highest frequency) of the involved functions, $dt' \sim 1/(4\pi F_\pi)$. Together with $A^\mu_{s}(t'n) \sim p$, we thus obtain an overall $p/4\pi F_\pi \sim \epsilon$ suppression. For the local case $t'=t$, standard chiral suppression applies.

All in all, the matching of the four-quark operator in SCET-2 in the heavy quark limit takes the form
\begin{align}
    O^{(1)}_{ijk}(s,t) &=\sum_{\alpha}  \int ds_1\cdots ds_a  \  C^{(1)}_{\alpha} (s,\dots, s_a,t,\mu) \times [J_{c,\alpha}^{(1)}(s,\dots, s_a)]_{ki} [J^{(1)}_{s}(t)]_j \, \, , 
\label{eq:full_op}
\end{align}
where the leading soft and collinear currents are defined in Eqs.~\eqref{eq:Jsdef} and~\eqref{eq:J2def}. As in the standard HMChPT, the nonperturbative matching coefficients, now distributed along the light cone, are to be related to a hadronic form factor by computing the relevant hadronic matrix element in the chiral limit.

In principle, this includes an infinite series of unknown coefficients, but these are effectively reduced to one, since an insertion of $J_{c,\alpha}^{(1)}$ will result in a vertex with emission of $\geq a$ collinear pions. In consequence, to describe the contribution of $O^{(1)}$ to $B \to \pi$ transitions at tree level in the hadronic theory, only the collinear current $J_{c,1}^{(1)}$ need be considered. In the chiral limit, 
\begin{align}
    \bra{\pi^a}O_{i}^{(1)}(s,t)\ket{\bar B_j}&=  \int ds_1 \  C^{(1)}_{1} (s,s_1,t,\mu) \times \bra{\pi^a} [J_{c,1}^{(1)}(s,s_1)]_{mi} [J^{(1)}_{s}(t)]_m  \ket{\bar B_j} \, ,\label{eq:O1ME}
\end{align}
where $O_{i} \equiv O_{imn} \delta_{mn}$. In the hadronic theory, the matrix element is equivalent to the product of separate matrix elements of the soft and the collinear side,
\begin{align}
    \bra{\pi^a} [J_{c,1}^{(1)}(s,s_1)]_{mi} [J^{(1)}_{s}(t)]_m  \ket{\bar B_j} &= \bra{\pi^a} [J_{c,1}^{(1)}(s,s_1)]_{mi} \ket{0} \times \bra{0} [J^{(1)}_{s}(t)]_m  \ket{\bar B_j} \, .
\end{align}
The equation above is simply a reflection of the factorization of soft and collinear fields at leading power in the SCET-2 \emph{Lagrangian}, and holds to all orders in the chiral expansion, after all necessary counterterms have been included. However, this is not a statement about the factorizability of the soft-overlap form factor; the hadronic coefficient in Eq.~\eqref{eq:O1ME} actually depends on the light-cone separations of both sectors in a non-trivial way. As already mentioned above, this can be traced back to the renormalization of the original 4-quark operators in SCET-2 which breaks factorization through the necessity to introduce additional messenger modes and/or regulators in the IR. The associated dynamical effects thus correspond to the UV sector of the hadronic theory, to be reflected in the non-perturbative matching coefficients for heavy-to-light currents. In contrast, for the SCET-2 operators describing the \emph{factorizable} contributions in Eq.~\eqref{eq:ffs_sum}, the soft and collinear operators renormalize independently, and, as a consequence, the non-perturbative matching coefficients of the associated soft-collinear hadronic currents factorize and can be identified with the LCDAs appearing in 
Eq.~\eqref{eq:ffs_sum}. 

The tree level matrix element then finally reduces to
\begin{align}
    \bra{\pi^a(p)}O_{i}^{(1)}(s,t)\ket{\bar B_j(v)} = -2\frac{\bar n\cdot p}{F_\pi}t^a_{ji}\int ds_1 \ C_{1}^{(1)}(s,s_1,t) \ e^{is_1\bar n\cdot p} \, .
\end{align}

The matching of the remaining operators $O^{(2-4)}$ to the hadronic theory and the evaluation of their tree-level matrix elements proceeds in parallel to the case of $O^{(1)}$. The operators $O^{(2)}$ and $O^{(3)}$ transform differently than $O^{(1)}$ in Eq.~\eqref{eq:Jctrans} because of the chirality of the valence quark, namely
\begin{align}
    J_s^{(2,3)} &\mapsto J_s^{(2,3)} R_s^{\dagger}(tn)\, , \quad J_c^{(2,3)} \mapsto R_c(s\bar n)J_c^{(2,3)} L_c^\dagger(0) \, . 
    \label{eq:O2trans}
\end{align}
Their collinear bilinears can be written as
\begin{align}
    \bar{\mathscr{X}}_{L,i}(0)\,\frac{\slashed{\bar n}}{2} \, i\slashed{D}_{\perp} \mathscr{X}_{R,k}(s\bar n) &= \frac{\bar n _\rho}{2}g^\perp_{\mu\nu} \bigl[\bar{\mathscr{X}}_{L,i}(0)\,\gamma^\rho\gamma^\mu \, iD^\nu \mathscr{X}_{R,k}(s\bar n)\bigr] \, ,\\
    \bar{\mathscr{X}}_{L,i}(0)\,\frac{\slashed{\bar n}}{2} \, \slashed{\mathscr{A}}_{c\perp}(r\bar{n}) \mathscr{X}_{R,k}(s\bar n) &= \frac{\bar n _\rho}{2}g^\perp_{\mu\nu} \bigl[\bar{\mathscr{X}}_{L,i}(0)\,\gamma^\rho \, \gamma^\mu \mathscr{A}^\nu_{c}(r\bar{n}) \mathscr{X}_{R,k}(s\bar n)\bigr] \, .
\end{align}
To build the equivalent hadronic current, the only leading open index structure that does not vanish when contracted with $\bar n _\rho \,g^\perp_{\mu\nu}$  is $\mathcal{A}^\rho_c \, g^{\mu\nu}$. In consequence, both operators simply match to
\begin{align}
    J_{c,1}^{(2)}(s,s_1) = \Xi_{c,R}^\dagger(s\bar n) \times \bar n\cdot \mathcal{A}_c (s_1\bar n)\times \Xi_{c,L}^\dagger(0)  \, ,
\end{align}
to be compared with Eq.~\eqref{eq:J1def}.\footnote{The contribution of the $J^{(2)}_{c,0}$ current to the $B \to \pi_c$ matrix element can be absorbed into $J_{c,1}^{(2)}$, as discussed in Appendix \ref{Appendix:J0_redundancy}.} Once again, the hadronic operator basis is extended to a general class of operators with an arbitrary number of insertions of the large component of the axial field
\begin{align}
    J_{c,\alpha}^{(2)}(s,s_1,\dots, s_a) &= \Xi_{c,R}^\dagger(s\bar n) \times \left[ \bar{n}\cdot \mathcal{A}_c(s_1\bar{n})\times \cdots \times \bar{n}\cdot \mathcal{A}_c(s_a\bar{n}) \right]\times \Xi_{c,L}^\dagger(0) \, ,
\end{align}
with the caveat that the Wilson coefficient for $O^{(3)}$ will inherit its $r$ dependence. Since the operator does not depend on $r$, we can perform the integral over $r$ in the hadronic representation of Eq.~\eqref{eq:QCD_to_SCETI}. Both $O^{(2)}$ and $O^{(3)}$ share the same soft current $\sim \bar{\mathscr{Q}}_{R,j}(tn)\slashed n \mathscr{H}_v(0)$, which carries the opposite chirality compared to $O^{(1)}$, and matches to the hadronic operator
\begin{align}
    J_{s}^{(2)}(t) &= \mathrm{Tr}[\mathcal{H}_v(0)\slashed{n} (1+\gamma_5)] \, \Xi_{s,R}(tn) \, .
\end{align}
Note that the trace above is the same as in Eq.~\eqref{eq:Jsdef}. The only collinear operator that contributes at tree level is $J_{c,1}^{(2)}$, so once again the tree level matrix elements reduce to
\begin{align}
    \bra{\pi^a(p)}O_{i}^{(2)}(s,t)\ket{\bar B_j(v)}_{m_q \to 0} &= -2\frac{\bar n\cdot p}{F_\pi}t^a_{ji}\int ds_1 \ C_{1}^{\textrm{(2)}}(s,s_1,t) \ e^{is_1\bar n\cdot p} \, , \\
    \bra{\pi^a(p)}O_{i}^{(3)}(s,r,t)\ket{\bar B_j(v)}_{m_q \to 0} &= -2\frac{\bar n\cdot p}{F_\pi}t^a_{ji}\int ds_1 \ C_{1}^{(3)}(s,s_1,r,t) \ e^{is_1\bar n\cdot p} \, .
\end{align}

Finally, $O^{(4)}$ transforms the same as $O^{(1)}$ in Eq.~\eqref{eq:Jctrans}. It has only a modified form of the soft current compared to $O^{(1)}$, which depends on the light-cone separation $un$. Once again, this will be inherited by the Wilson coefficient, but the hadronic operator will not depend on $u$. With these considerations in mind, the whole operator ends up matching to the same hadronic operators as $O^{(1)}$, so the tree level matrix element is once again
\begin{align}
    \bra{\pi^a(p)}O_{i}^{(4)}(s,t,u)\ket{\bar B_j(v)}_{m_q \to 0} = -2\frac{\bar n\cdot p}{F_\pi}t^a_{ji}\int ds_1 \ C_{1}^{(4)}(s,s_1,t,u) \ e^{is_1\bar n\cdot p} \, .
\end{align}

To relate these matrix elements to the soft overlap form factor, and to later state their chiral corrections, it is useful to define, including the perturbative ingredients from the partonic matching onto SCET-2 (recovering the renormalization scale dependence which was left implicit above),
\begin{align}
    \zeta_{\pi,L}(E_\pi,\mu) &\equiv \zeta_{\pi}^{(1)}(E_\pi,\mu) + \zeta_{\pi}^{(4)}(E_\pi,\mu) \, ,\\
    \zeta_{\pi,R}(E_\pi,\mu) &\equiv \zeta_{\pi}^{(2)}(E_\pi,\mu) + \zeta_{\pi}^{(3)}(E_\pi,\mu) \, ,
\end{align}
where for each operator we can define their contribution by including the perturbative kernel $D^{(\gamma)}$ as well, as an example
\begin{align}
    \zeta_{\pi}^{(1)}(E_\pi,\mu)_{m_q \to 0} &\equiv -\frac{2}{F_\pi}\int ds\,dt\; D^{(1)}(s,t,\bar{n}\cdot p,\mu) \int ds_1 \,  C_{1}^{(1)}(s,s_1,t,\mu) \ e^{is_1\bar{n}\cdot p} \nonumber \\
&=  -\frac{2}{F_\pi} (\bar n \cdot p) \int_0^1 \frac{du}{2\pi} \int_0^\infty \frac{d\omega}{2\pi}  \widetilde{D}^{(1)}(u, \omega, \bar{n}\cdot p, \mu) \ \widetilde{C}_{1}^{(1)}(u,\omega, \bar{n}\cdot p,\mu) \, , \label{eq:ffmom}
\end{align}
with $\bar{n}\cdot p=2E_\pi$ and
\begin{align}
\widetilde{D}^{(1)}(u,\omega, \bar{n}\cdot p,\mu) &= \int ds \, dt \  e^{ius \bar{n}\cdot p}e^{-i\omega t}\, D^{(1)}(s,t,\bar{n}\cdot p,\mu) \, , \\
\widetilde{C}_{1}^{(1)}(u,\omega, \bar{n}\cdot p,\mu) &= \int ds \, dt \, ds_1 \  e^{-ius\bar{n}\cdot p}e^{i\omega t} e^{is_1\bar{n}\cdot p}\,C_{1}^{(1)}(s,s_1,t,\mu) \, .
\end{align}
The support of the integrals in Eq.~\eqref{eq:ffmom} is enforced by the perturbative kernel. We have grouped $O^{(1)}$ with $O^{(4)}$ and $O^{(2)}$ with $O^{(3)}$ since they match to the same hadronic operators. Finally, the sum of the two groups 
is just the usual soft-overlap form factor
\begin{align}
    \zeta_{\pi}(E_\pi,\mu)=\zeta_{\pi,L}(E_\pi,\mu)+\zeta_{\pi,R}(E_\pi,\mu) \, .
\end{align}
With this result we achieved our first goal, finding the representation of the soft-overlap $B\to \pi$ form factor within the hadronic soft-collinear effective theory.


\section{Chiral logarithms}
\label{sec:4}


\subsection{Soft-overlap form factor}

Having established the essential structure of the effective theory, we now study the one-loop corrections to the hadronic representation of the soft-overlap currents derived in the previous section. As usual in the ChPT framework, these contribute terms of order $m_\pi^2 \ln m_\pi^2$, which are also attributed to the renormalization of sub-leading operators (the coefficients of the latter also provide analytic corrections to the chiral limit of order $m_\pi^2$ without logarithmic enhancement).

\begin{figure}[t]
  \centering

  \begin{minipage}[b]{0.25\textwidth}
    \centering
    \includegraphics[width=\linewidth]
      {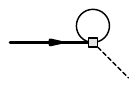}\\[-1mm]
    (a)
  \end{minipage}
  \hfill
  \begin{minipage}[b]{0.25\textwidth}
    \centering
    \includegraphics[width=\linewidth]
      {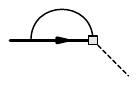}\\[-1mm]
    (b)
  \end{minipage}
  \hfill
  \begin{minipage}[b]{0.25\textwidth}
    \centering
    \includegraphics[width=\linewidth]
      {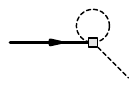}\\[-1mm]
    (c)
  \end{minipage}

  \caption{Nonvanishing corrections to the $B\to\pi_c$ matrix element (at low-$q^2$) of the hadronic realization of a non-local four-quark operator, depicted here for simplicity as a single vertex. The heavy-hadron multiplet, soft pions, and collinear pions are represented by the solid arrow, solid lines, and dashed lines, respectively. Only the tadpole topologies~(a) and~(c) give rise to chiral-logarithmically
  enhanced contributions.}
  \label{fig:B_loops}
\end{figure}

The one-loop $B \to \pi$ matrix elements of hadronic currents can be categorized into `tadpole' and `soft-exchange' terms, shown in Fig.~\ref{fig:B_loops}. To calculate the tadpole-type contributions, one has to consider every possible expansion of the fields in the non-local effective vertex that results in two soft pion fields and one collinear pion field, or three collinear pion fields. In the former case, the two soft pion fields are contracted in a loop, and in the latter, any two of the collinear fields are contracted.  The presence of the additional Wilson lines makes the combinatorics somewhat involved. However, many individual contractions vanish, due to group-algebraic identities (such as $f^{abc} \delta^{ab}=0$), trivial tensor reduction of momentum integrals, and/or light-cone vector identities. Notably, all contractions involving pions originating from a Wilson line vanish in the considered matrix elements, as $\bar{n}^2=0$.\footnote{Interestingly, for $J^{(\gamma)}_{c,0}$ the Wilson lines \emph{do} contribute to the chiral logarithms, unlike for $J^{(\gamma)}_{c,1}$: one pion from the expansion of the Wilson line is emitted as the external state, while a second contracts with a first-order endpoint field~$\xi$. We explicitly verified that the coefficient of the chiral logarithm is the same as for $J^{(\gamma)}_{c,1}$ nonetheless, as anticipated by the matrix-element relations of Appendix~\ref{Appendix:J0_redundancy}.}

At first glance, the collinear (or soft) tadpole integrals we encounter have a dependence on the non-locality $s$ (or $t$),
\begin{align}
    I_1(s) \equiv -i \tilde{\mu}^{2\epsilon}\int\frac{d^dk}{(2\pi)^d} \frac{e^{-is\bar n\cdot k}}{k^2-m_\pi^2}  \, ,
    \label{eq:tad_s}
\end{align}
where $\tilde{\mu}^2=\mu^2 e^{\gamma_E} /(4\pi)$. However, the integrals simply evaluate to the standard one-point integral because of boost invariance
\begin{align}
    I_1(s)=I_1(0) &=  \frac{m_\pi^2}{(4\pi)^{2}} \left( \frac{1}{\epsilon} - \ln \frac{m_\pi^2}{\mu^2} + 1 \right) \, .
\end{align}
Contractions of the form of Fig.~\ref{fig:B_loops}(b) reduce to the integral
\begin{align}
    I_2(t) &= n^\nu (g_{\nu\mu}-v_\nu v_\mu) \tilde{\mu}^{2\epsilon}\int \frac{d^dk}{(2\pi)^d}\frac{e^{-itn\cdot k}}{k^2-m_\pi^2}  \frac{k^\mu}{v\cdot k+ \Delta} \, . \label{eq:I2}
\end{align}
For $t=0$ (corresponding to local HMChPT), the integral is proportional to $v^\mu$, which vanishes when contracted with the vector meson propagator. We show in Appendix \ref{Appendix:no_chiral_logs} that $I_2(t,m_\pi^2)$ is analytic in $m_\pi^2$ for $t\neq 0$ (to the order considered), and in particular does not depend on either of the two logarithmically enhanced structures $\Delta^2 \ln m_\pi^2$ or $m_\pi^2 \ln m_\pi^2$. Therefore we can safely neglect the two-point integral topology, as far as chiral logarithms are concerned. We can now see that all our soft currents receive the same logarithmically enhanced chiral corrections. They only differ in non-logarithmically-enhanced terms, from the $B^*$ pole loop, and unspecified values of the sub-leading low-energy constants.

\begin{figure}[t]
  \centering

  \begin{minipage}[b]{0.20\textwidth}
    \centering
    \includegraphics[width=\linewidth]
      {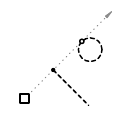}\\[-1mm]
    (a)
  \end{minipage}
  \hfill
  \begin{minipage}[b]{0.20\textwidth}
    \centering
    \includegraphics[width=\linewidth]
      {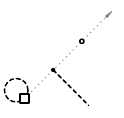}\\[-1mm]
    (b)
  \end{minipage}
  \hfill
  \begin{minipage}[b]{0.20\textwidth}
    \centering
    \includegraphics[width=\linewidth]
      {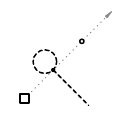}\\[-1mm]
    (c)
  \end{minipage}
  \hfill
  \begin{minipage}[b]{0.20\textwidth}
    \centering
    \includegraphics[width=\linewidth]
      {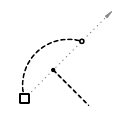}\\[-1mm]
    (d)
  \end{minipage}

  \caption{Detailed expansion of the collinear pion loops shown collectively in Fig.~\ref{fig:B_loops}(c), outlining the hadronic fields defined on the $\bar{n}$-ray (shown as a light gray arrow). The coordinates $s\bar{n}$ and $s_1\bar{n}$ are indicated by the open circle and solid dot, respectively (also as in Fig.~\ref{fig:combined}).}
  \label{fig:B_loops_wilson}
\end{figure}

As we mentioned before, currents with two or more $\bar{n}\cdot \mathcal{A}_c$ insertions (see Eq.~\eqref{eq:J2def}) are not chirally suppressed, but do not contribute to $B \to \pi_c$ at tree level. At one-loop level then, for the currents with two $\bar{n}\cdot \mathcal{A}_c$ insertions, there are two possible Wick contractions, either with one derivative or two. For the currents with three insertions, the only possible contraction contains two derivatives. These cases result in the following tadpole-type integrals
\begin{align}
    -i\bar n ^\mu \int\frac{d^d k}{(2\pi)^d} \ e^{is\bar n\cdot k}\frac{k_\mu}{k^2-m_\pi^2} \, , \qquad
    -i \bar n ^\mu\bar n^\nu \int\frac{d^dk}{(2\pi)^d} \ e^{is\bar n\cdot k}\frac{k_\mu k_\nu}{k^2-m_\pi^2}  \, ,
\end{align}
which both vanish as $\bar n^2=0$. Finally, it is straightforward to show that currents with more than three insertions would contribute at least at two-loops.

Due to the form of the operator, it is informative to separate the one-loop corrections of the soft and collinear sectors. We define
\begin{align}
    L_\pi &\equiv \frac{m_\pi^2}{(4\pi F_\pi)^{2}} \ln \frac{\mu^2}{m_\pi^2} \, . \label{eq:L_pi}
\end{align}
The one-loop logarithmically enhanced corrections to the collinear matrix element are given by
\begin{align}
    \bra{\pi^a}J_{c,1}^{(1)}(s,s_1) \ket{0} &=  \sqrt{R_{\pi}} \ \bra{\pi^a}J_{c,1}^{(1)}(s,s_1) \ket{0} _{m_q \to 0} \, \Big[1+\frac{2C_A}{3}  L_\pi \Big] \, , \\
    \bra{\pi^a}J_{c,1}^{(2)}(s,s_1) \ket{0} &=  \sqrt{R_{\pi}} \ \bra{\pi^a}J_{c,1}^{(2)}(s,s_1) \ket{0} _{m_q \to 0} \, \Big[1+ \Big( 2C_F-\frac{C_A}{3}\Big) L_\pi \Big] \, ,
\end{align}
where $C_A$ and $C_F$ are the usual quadratic Casimir invariants of $SU(N)$. We have implicitly absorbed the UV divergences of the matrix elements into the renormalization of higher-order current operators that include mass-spurion insertions $\mathcal{S}_a$ as defined in Eq.~\eqref{eq:dressed}. The collinear pion external-leg corrections are equivalent to those of standard ChPT
\begin{align}
    \sqrt{R_{\pi}} &= 1 - \frac{C_A}{6} L_\pi + \textrm{finite} \, .
\end{align}

Interestingly, the result for $J_{c,1}^{(1)}$ is the same as the correction to the pion decay constant calculated in standard ChPT \cite{Gasser:1984gg}. This is due to the fact that the tadpole integral that appears in Eq.~\eqref{eq:tad_s} actually reduces to the local equivalent, and the Wilson line contributions vanish. Indeed, if we consider the local limit of our operator, it reduces to the usual current that defines the pion decay constant
\begin{align}
    J_{c,1}^{(1)}(0,0)=[\Xi_{c,L}\, \bar{n}\cdot \mathcal{A}_c \, \Xi_{c,L}^\dagger](0) = [\xi_c \ \bar n \cdot A_c  \ \xi_c^\dagger] (0) = -\bar n_\mu  \frac{i}{2}[\Sigma_cD^\mu \Sigma_c^\dagger](0) \, .
\end{align}
On the other hand, the corrections to the collinear current $J_{c,1}^{(2)}$ are different. Namely, one of the tadpole contributions, the one depicted in Fig.~\ref{fig:B_loops_wilson}~(d), receives a relative minus sign. If we now take the local limit of $J_{c,1}^{(2)}$
\begin{align}
    J_{c,1}^{(2)}(0,0)=[\Xi^\dagger_{c,R} \, \bar{n}\cdot \mathcal{A}_c \, \Xi_{c,L}^\dagger](0) = [\xi^\dagger_c \ \bar n \cdot A_c  \ \xi_c^\dagger] (0)=-\bar n_\mu  \frac{i}{2}[D^\mu \Sigma_c^\dagger](0) \, ,
    \label{eq:local_mu_pi}
\end{align}
which corresponds to the local current that is used in the normalization of higher twist pion LCDAs \cite{Chen:2005js}.

For the case of the soft operators, our result reads
\begin{align}
    \bra{0} J_{s}^{(\gamma)}(t)\ket{\bar B} =  \sqrt{R_B} \ \bra{0} J_{s}^{(\gamma)}(t)\ket{\bar B}_{m_q \to 0} \, \Big[1+ \frac{C_F}{2}L_\pi \Big] \, ,
\end{align}
where the $B$-meson external leg corrections are again equivalent to those of standard HMChPT
\begin{align}
    \sqrt{R_B} &= 1 + \frac{3C_F}{2}g_\pi^2 \left(1 - \frac{2\Delta^2}{m_\pi^2}\right) L_\pi +\textrm{finite}\ \, .
\end{align}
Once again, this coincides with the $B$-meson decay constant chiral corrections \cite{Goity:1992tp,Grinstein:1992qt, Boyd:1994pa}, again due to the fact that the Wilson lines do not contribute at this level, and that the non-local version of the soft-exchange does not contribute chiral logarithms.

Since the chiral corrections are multiplicative and independent of the space-time positions, the sum of all the chiral-logarithmically enhanced one-loop contributions can be written as
\begin{align}
    \zeta_{\pi,L}(E_\pi,\mu) &= \zeta_{\pi,L}(E_\pi,\mu)_{m_q \to 0} \left[ 1+c_{\pi,L} L_\pi\right] \, ,\\
    \zeta_{\pi,R}(E_\pi,\mu) &= \zeta_{\pi,R}(E_\pi,\mu)_{m_q \to 0} \left[ 1+c_{\pi,R} L_\pi\right] \, ,
\end{align}
where the coefficients of the logarithms, specified to the case $N=2$ ($C_F=3/4$, $C_A=2$) are given by
\begin{align}
    c_{\pi,L} = \frac{C_A}{2} + \frac{C_F}{2}\left[1+3g_\pi^2\left(1-\frac{2\Delta^2}{m_\pi^2}\right)\right] &=\frac{11}{8}+\frac{9}{8}g_\pi^2\left(1-\frac{2\Delta^2}{m_\pi^2}\right) \, ,\label{eq:c_L} \\
    c_{\pi,R} = 2C_F-\frac{C_A}{2} + \frac{C_F}{2}\left[1+3g_\pi^2\left(1-\frac{2\Delta^2}{m_\pi^2}\right)\right] &= \frac{7}{8}+\frac{9}{8}g_\pi^2\left(1-\frac{2\Delta^2}{m_\pi^2}\right) \, . \label{eq:c_R}
\end{align}

Since the coefficients $c_\pi$ are different for both pieces, and the relationship between the non-perturbative objects $\zeta_{\pi,L}$ and $\zeta_{\pi,R}$ is not known, the chiral corrections to the full soft overlap form factor cannot be written as an overall multiplicative term, namely
\begin{align}
    \zeta_{\pi}(E_\pi,\mu) &\neq \zeta_{\pi}(E_\pi,\mu)_{m_q \to 0} \left[ 1+c_{\pi} L_\pi\right].
\end{align}
Instead, we generically have
\begin{align}
    \zeta_{\pi}(E_\pi,\mu) &= \zeta_{\pi,L}(E_\pi,\mu)_{m_q \to 0} \left[ 1+c_{\pi,L} L_\pi\right]+\zeta_{\pi,R}(E_\pi,\mu)_{m_q \to 0} \left[ 1+c_{\pi,R} L_\pi\right] \, .
\end{align}
Therefore the chiral extrapolation of the $B \to \pi$ soft overlap form factor depends on unknown non-perturbative weights and therefore cannot be established within the effective theory.


\subsection{LCDAs and the factorizable contribution}


Having analyzed the non-factorizable part of the form factors, one can then come back to the case of the factorizable part. The leading twist LCDAs that enter in Eq.~\eqref{eq:ffs_sum} are defined as
\begin{align}
\bra{\pi^a(p)} \bar{\mathscr{X}}_{L,i}(0)\,\slashed{\bar n}\,\mathscr{X}_{L,j}(s\bar n) \ket{0}
&= -i F_\pi \,(\bar n \cdot p)\int_0^1 du\, e^{ius\bar n\cdot p}\,\phi_{\pi}(u,\mu) \, t^a_{ji}\,, \\
\bra{0}\,\bar{\mathscr{Q}}_{L,i}(tn)\,\slashed{n}\,\mathscr{H}_v(0)\ket{\bar B_j(v)}
&= i\,\frac{F_B(\mu)}{2}\int_0^\infty d\omega\,e^{-i\omega t}\,\phi_B^+(\omega,\mu)\, \delta_{ji}.
\end{align}
The matching of the quark bilinears is parallel to that of the four-quark operators. Namely, we find that at tree and one-loop level
\begin{align}
    \bra{\pi^a(p)} \bar{\mathscr{X}}_{L,i}(0)\,\slashed{\bar n}\,\mathscr{X}_{L,j}(s\bar n) \ket{0} &= \int ds_1 \, C_\pi(s,s_1,\mu) \bra{\pi^a(p)} [J_{c,1}^{(1)}(s,s_1)]_{ji}\ket{0} \, ,\\
    \bra{0}\bar{\mathscr{Q}}_{L,i}(tn)\,\slashed{n}\,\mathscr{H}_v(0)\ket{\bar B_j(v)} &= C_{B} (t,\mu) \bra{0} [J_{s}^{(\text{LCDA})}(t)]_i\ket{\bar B_j(v)} \, ,
\end{align}
with
\begin{align}
    J_{s}^{(\text{LCDA})}(t) =\Tr[\mathcal H_v(0)\slashed n (1-\gamma_5)] \, \Xi_{s,L}^\dagger(tn) \, ,
\end{align}
which differs from $J_{s}^{(1)}$ in Eq.~\eqref{eq:Jsdef} by the projector that enters the trace.

We can now relate the hadronic constants to the LCDAs. Evaluating the matrix elements gives
\begin{align}
    \bra{\pi^a(p)} [J_{c,1}^{(1)}(s,s_1)]_{ji}\ket{0}_{m_q=0} &= -i\frac{\bar n\cdot p}{F_\pi}e^{is_1\bar n\cdot p} \ t^a_{ji} \, ,\\
    \bra{0} [J_{s}^{(\text{LCDA})}(t)]_i\ket{\bar B_j(v)}_{m_q=0} &= 2i \delta_{ji}\, .
\end{align}
Using this result, we now get
\begin{align}
    \int ds_1\,C_\pi(s,s_1,\mu)e^{is_1\bar n \cdot p} &= F_\pi^2 \int_0^1du\,e^{iu s\bar n \cdot p}\phi_\pi(u,\mu) \, , \\
    2 \, C_{B}(t,\mu) &= \frac{F_B(\mu)}{2}\int_0^\infty d\omega\,e^{-i\omega t}\,\phi_B^+(\omega,\mu)\, .
\end{align}
One obtains that the LCDAs in the chiral limit are momentum space representations of the hadronic coefficients
\begin{align}
    \phi_\pi(u,\mu)&=\frac{\bar n \cdot p}{F_\pi^2} \int \frac{ds}{2\pi}\,e^{-iu s\bar n \cdot p} \int ds_1\,C_\pi(s,s_1,\mu)e^{is_1\bar n \cdot p} \, , \\
    \phi_B^+(\omega,\mu) &= \frac{4}{F_B(\mu)}\int \frac{dt}{2\pi}\, e^{i\omega t}C_{B}(t,\mu) \, .
\end{align}

As we already saw in the previous section, the logarithmically-enhanced chiral corrections of these hadronic operators are the same as that of their respective decay constants. In consequence, at this order, the LCDAs do not receive such chiral corrections. The chiral logarithm is carried only by the decay constants, which is consistent, for instance, with Refs.~\cite{Chen:2003fp,Grinstein:2005ry}. Furthermore, we can calculate the chiral corrections to any higher-twist LCDA by matching the corresponding partonic matrix element to a non-local hadronic operator. Additionally, since any LCDA moment can be represented as the matrix element of the corresponding smeared light-ray operator, it would be described by the same hadronic operator as the original LCDA and therefore receive the same one-loop chiral corrections. This implies that the leading chiral corrections to the normalized moments of the distribution amplitudes (namely, the Gegenbauer moments, see for instance Refs.~\cite{Ball:2006wn,Braun:2006dg}) are not logarithmically enhanced. Our work provides a theoretical basis for the procedure of neglecting $SU(3)$ breaking of the Gegenbauer moments while retaining $SU(3)$ breaking in the decay constants and form factors in recent applications to nonleptonic $B \to PP$ decays within QCD factorization~\cite{BurgosMarcos:2025xja, Fang:2026fhl}.

As for how this enters the full form factor, in the same language as we were using for the soft-overlap form factor, we would be matching the operator
\begin{align}
    O_{ijk}^{(\text{LCDA})} &= \,\bigl[\bar{\mathscr{X}}_{L,i}(0)\,\frac{\slashed{\bar n}}{2}\,\mathscr{X}_{L,k}(s\bar n)\bigr]\,
   \bigl[\bar{\mathscr{Q}}_{L,j}(tn)\,\frac{\slashed n}{2} \,\mathscr{H}_v(0)\bigr] \, .
\end{align}
The only difference between this operator and $O^{(1)}$ is the soft Dirac structure, and consequently matches to the same collinear hadronic operators and receives the same chiral corrections. Their contribution can then be effectively absorbed into the unknown function $f_{i,L}^{B \to \pi}(E_\pi)$ for the purpose of computing chiral corrections for the form factor. The left-handed chirality contribution to the form factors consists of both factorizable and non-factorizable terms
\begin{align}
    f_{i,L}^{B \to \pi}(E_\pi) &\equiv C_i(E_\pi,\mu)\zeta_{\pi,L}(E_\pi,\mu) \nonumber\\
    &\hspace{1cm}+ \int_0^\infty d\omega \int_0^1 du\;
    F_B(\mu) \, \phi_B^+(\omega,\mu)\,
    T_i(E_\pi,\omega,u,\mu)\,
    F_\pi \, \phi_\pi(u,\mu) \, ,\label{eq:f_L}
\end{align}
whereas the right-handed sector remains purely non-factorizable,
\begin{align}
     f_{i,R}^{B \to \pi}(E_\pi) &\equiv C_i(E_\pi,\mu)\zeta_{\pi,R}(E_\pi,\mu) \, .\label{eq:f_R}
\end{align}
In the very same way as we did for the soft overlap form factor, the chiral corrections for the vector and scalar form factors $f_{+/0}$ are written as
\begin{align}
    f_{i}^{B \to \pi}(E_\pi) &\simeq f_{i,L}^{B \to \pi}(E_\pi)_{m_q \to 0} \left[ 1+c_{\pi,L} L_\pi\right]+f_{i,R}^{B \to \pi}(E_\pi)_{m_q \to 0} \left[ 1+c_{\pi,R} L_\pi\right] \, ,
    \label{eq:f_result}
\end{align}
with the same chiral log coefficients defined in Eqs. \eqref{eq:c_L} and  \eqref{eq:c_R}.

After this exercise one might be confused about the difference between the factorizable and non-factorizable hadronic matching. The main difference is the position dependent Wilson coefficients that appear in each case. In the case of the factorizable contribution, the hadronic coefficient can be written as a product of two hadronic coefficients that depend either on the nonlocality in the collinear direction (the pion distribution amplitude) and the non-locality associated to soft fluctuations along the light-cone (the $B$ meson distribution amplitude).
\begin{align}
     C_{1}^{(\gamma)}(s,s_1,t) &\neq C(s,s_1) \cdot C(t) \, , \\
     C^{(\text{LCDA})}(s,s_1,t) &= \frac{1}{4}C_\pi(s,s_1) \cdot C_{B}(t) \, .
\end{align}
So no new information about the factorizability of the two contributions is claimed, just that the same approach to calculate chiral logarithms can be used in both cases.


\subsection{Comparison to low-recoil and Hard-Pion ChPT}
\label{sec:4.3}


It is worthwhile to compare our results for the coefficients of chiral logarithms at low-$q^2$ to the analogous results in HMChPT at high-$q^2$. At low-$q^2$, we have shown that the soft contributions to the form factors $f_+$ and $f_0$ are equivalent in the heavy quark limit and their chiral corrections are accordingly the same, although they originate from two different chirality sectors with independent energy dependence. In contrast, at high-$q^2$, the $f_+$ form factor is strongly affected by the nearby $B^*$ meson, such that the hyperfine splitting $\Delta = M_{B^*}-M_B \sim O(\Lambda^2/m_b)$ must be retained, and the vector and scalar form factors are unrelated. One can write the chiral corrections at low recoil as
\begin{align}
    f_i(q^2) &\simeq \left.f_i(q^2)\right|_{m_q \to 0} [ 1 + c_i(q^2) L_\pi ] \, , \qquad (q^2\sim m_b^2) \, ,
\end{align}
where $L_\pi$ is given in Eq.~\eqref{eq:L_pi}. 
For general number $N$ of mass-degenerate flavors, we find
\begin{align}
    c_0(q^2) &=\frac{C_F}{2}-\frac{C_A}{2} +C_A\,\frac{E_\pi^2}{m_\pi^2} +3C_F\,g_\pi^2\left(\frac{1}{2}-\frac{\Delta^2}{m_\pi^2}\right), \\
    c_+(q^2) &=\frac{C_F}{2} +g_\pi^2\Bigg[ \left(\frac{11C_F}{2}-\frac{C_A}{2}\right) +\left(\frac{C_A}{3}-\frac{8C_F}{3}\right)\frac{E_\pi^2}{m_\pi^2} +\left(\frac{5C_A}{3}-6C_F\right)\frac{E_\pi\Delta}{m_\pi^2} \nonumber \\ &\hspace{2.2cm} +\left(\frac{5C_A}{3}-\frac{23C_F}{3}\right)\frac{\Delta^2}{m_\pi^2} -C_F\,\frac{\Delta}{E_\pi+\Delta}\left(1-\frac{2}{3}\frac{E_\pi^2}{m_\pi^2}\right) \Bigg].
    \label{eq:c_plus}
\end{align}
The coefficients increase quadratically with $E_\pi$, signaling the breakdown of the chiral expansion in HMChPT at large energy. Note that the last term in Eq.~\eqref{eq:c_plus} includes the expansion of the $\Delta$ parameter in the tree level expression for the vector form factor around its chiral limit. 

In the limit $\Delta\to 0$ we reproduce the result of Ref.~\cite{Becirevic:2003ad} for general $N$. The $SU(2)$ and $SU(3)$ special cases agree with the results in Refs.~\cite{Fleischer:1992tn,Falk:1993fr,Bijnens:2010ws} when the appropriate limits are taken. The full $\Delta$‑dependent terms are, to our knowledge, new. In the case $N=2$, at the endpoint $E_\pi = m_\pi$, the coefficients above simplify to
\begin{align}
    c_0(q^2_{\textrm{max}}) &= \frac{11}{8} + \frac{9}{8}g_\pi^2\left(1-\frac{2\Delta^2}{m_\pi^2}\right) \, , \\
    c_+(q^2_{\textrm{max}}) &= \frac{3}{8} + g_\pi^2 \left( \frac{43}{24} -\frac{7}{6}\frac{\Delta}{m_\pi} -\frac{29}{12}\frac{\Delta^2}{m_\pi^2} -\frac{1}{4}\frac{\Delta}{m_\pi+\Delta} \right) \, .
\end{align}
Curiously, the result for the coefficient of the scalar form factor $c_0(q^2_{\textrm{max}})$ coincides with our result for the left-handed spectator chirality sector $c_{\pi,L}$ at low-$q^2$ given in Eq.~\eqref{eq:c_L}. 
This equivalence is not expected or required, as the HMChPT computation has fundamentally different loop contributions.

Another way to understand the qualitative difference in the structure of the chiral corrections in different $q^2$ regions is to naively apply the four-quark operators appropriate to the low-$q^2$ region to the high-$q^2$ region. If we simply set $c=s$, while also localizing all of our operators, both collinear operators arising from the chirality sectors $\chi=L,R$ would collapse following $\xi^\dagger \xi = 1$ to the one soft operator.
Schematically,
\begin{gather} 
\bigg[ H_v \, \underbrace{[\cdots] \, \xi_{s,\chi}^\dagger \bigg](t_i) \times C(s_i,t_i) \times \bigg[ \xi_{c,\chi} \, [\cdots] } \, \xi_{c,L}^{\dagger}\bigg](s_i) 
\cr
 \updownarrow
\cr 
\bigg[ H_v \, [ \cdots ] \, \xi_{s,L}^\dagger\bigg](0)
\,. 
\end{gather}
Furthermore, the infinite tower of leading operators (implicit in the $[\cdots]$ above) would return to being suppressed by the derivative expansion, leaving only one leading operator $\sim H_v \xi^\dagger$.

\begin{figure}[t]
  \centering

  \begin{minipage}[b]{0.15\textwidth}
    \centering
    \includegraphics[width=\linewidth]
      {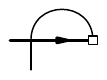}\\[-1mm]
    (a)
  \end{minipage}
  \hfill
  \begin{minipage}[b]{0.15\textwidth}
    \centering
    \includegraphics[width=\linewidth]
      {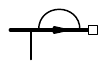}\\[-1mm]
    (b)
  \end{minipage}
  \hfill
  \begin{minipage}[b]{0.15\textwidth}
    \centering
    \includegraphics[width=\linewidth]
      {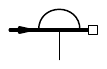}\\[-1mm]
    (c)
  \end{minipage}
  \hfill
  \begin{minipage}[b]{0.15\textwidth}
    \centering
    \includegraphics[width=\linewidth]
      {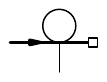}\\[-1mm]
    (d)
  \end{minipage}
  \hfill
  \begin{minipage}[b]{0.15\textwidth}
    \centering
    \includegraphics[width=\linewidth]
      {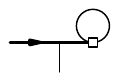}\\[-1mm]
    (e)
  \end{minipage}
  \hfill
  \begin{minipage}[b]{0.15\textwidth}
    \centering
    \includegraphics[width=\linewidth]
      {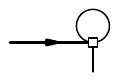}\\[-1mm]
    (f)
  \end{minipage}

  \caption{Nonvanishing corrections to the $B \to \pi_s$ matrix element (at high-$q^2$) in HMChPT. The square vertex indicates the hadronic realization of the local heavy-to-light current in HQET.
  See Fig.~\ref{fig:B_loops} for further details.}
  \label{fig:high-q2-loops}
\end{figure}

As already mentioned in the introduction, another approach to the computation of chiral logarithmic corrections to the heavy-to-light form factors at low $q^2$ is that of using the so-called Hard-Pion ChPT introduced in Ref.~\cite{Bijnens:2010ws}. The authors found that the chiral corrections to both form factors at low-$q^2$ and in the limit $\Delta=0$ can be written as
\begin{align}
    f_{i}(q^2)=f_{i}(q^2)_{m_q \to 0}\left [ 1+   \left(\frac{3}{8} + \frac{9}{8} g_\pi^2 \right) L_\pi \right] \, , \qquad \textrm{(HPChPT)}
\end{align}
in terms of a universal multiplicative factor. This is manifestly different from our results in Eq.~\eqref{eq:f_result}: while the term proportional to $g_\pi^2$ is given by the renormalization of the $B$-meson field in both cases, the analogue of the universal constant term $3/8$, in our case is to be replaced by two independent and different terms appearing in $c_{\pi,L}$ and $c_{\pi,R}$, respectively.

In their paper, the authors of \cite{Bijnens:2010ws} do not introduce explicit collinear dynamical modes (and do not consider their corresponding virtual corrections). Instead, they assume that the non-analytic chiral dependence is generated by soft-pion loops only, and that the effects of energetic pions are `hard' and can be absorbed into local effective couplings. Indeed, if in our framework we neglected collinear loops whatsoever, including the renormalization of the collinear pion, we could reproduce their result. However, the actual calculation in Ref.~\cite{Bijnens:2010ws} leads to a different set of chiral loop diagrams, which makes a direct comparison between the two approaches difficult. In our formalism, we treat both soft and collinear sectors as dynamical, and the resulting virtual corrections naturally appear from the matching of SCET-2 onto a hadronic theory. 

It is also to be stressed that the direct emission of an energetic pion from a heavy $B$-meson would result in a $b$-flavored state with invariant mass $(M_B v - E_\pi n)^2 \simeq q^2 \ll M_{B^*}^2$ which does not correspond to an effective hadronic degree of freedom in the EFT. Instead, the corresponding off-shell effects are encoded in the coefficient (functions) appearing in the SCET construction. Given that our analysis follows from matching SCET-2 including all relevant degrees of freedom, and we obtain a contradicting result, the approach of Ref.~\cite{Bijnens:2010ws} should probably be critically reconsidered before using it to guide extrapolations of lattice form factors at low-$q^2$.

\section{Summary}
\label{sec:5}

We have developed a non-local effective hadronic field theory for the $B \to \pi$ semileptonic form factors at large recoil, unifying the principles of Soft-Collinear Effective Theory (SCET) and Heavy-Meson Chiral Perturbation Theory (HMChPT) within a single framework. The essential development is a hadronic realization of non-local light-cone operators, organized within a large-energy/heavy-quark and chiral expansion. To reconcile the non-local structure inherited from SCET with chiral symmetry, we have introduced `chiral Wilson lines’ that connect spatially separated hadronic fields while preserving covariance under local chiral transformations.

Within this framework, we matched the four-quark operators entering the \mbox{SCET-2} description of the soft, non-factorizable heavy-to-light form factor onto their hadronic representations and determined their leading non-analytic dependence on the light quark/meson masses. Notably, the hadronic matching procedure bypasses the well-known (technical) problem of separating  the soft and collinear quark and gluon modes for exclusive $b$-quark decays in SCET. Instead, the non-factorizing nature of soft-collinear dynamics is reflected in the non-trivial dependence of the (unknown) hadronic matching coefficients on the light-cone separations of soft and collinear pion fields. 

In the chiral limit, the resulting soft $B\to\pi$ form factor is given as the matrix element of two independent sectors of hadronic operators associated to the chiral realizations of the SCET-2 operators. The two sectors are essentially related to the chirality of the spectator quark of the $B$ meson which is accelerated into the final state pion by strong interactions, and receive different chiral corrections. The origin of the distinct chiral logarithms is related to the leading and subleading distribution amplitudes that appear in the (naive) application of the QCD factorization approach to the soft form factor.

The same construction also applies to the non-local operators defining the light-cone distribution amplitudes (LCDAs) that appear in the factorizing form-factor contributions. Combining the factorizable and non-factorizable ingredients, we obtained the chiral corrections to the $B \to \pi$ form factors at low-$q^2$. In contrast to the low-recoil result, these corrections cannot be represented by a universal factor multiplying the full form factor in the chiral limit.

Our results are not consistent with those obtained in Hard-Pion Chiral Perturbation Theory (HPChPT), both qualitatively and quantitatively. In particular, HPChPT does not reproduce two distinct chiral corrections associated with the independent hadronic functions appearing in the chiral limit. In addition, even for the `leading' chirality sector with a left handed valence quark, we obtain a different chiral logarithm than in HPChPT. Part of the difference may be due to the fact that loop corrections in HPChPT are always associated with soft pions, 
whereas in our approach we argue that virtual corrections involving energetic pions are to be considered as well. This is also supported by our results for the factorizable form-factor contributions, where the logarithmically-enhanced chiral corrections are fully encoded in the renormalization of the corresponding meson decay constants in the soft \emph{and} collinear sector.

This disagreement between our results and those of HPChPT indicates that the form of the chiral extrapolation at large recoil is not yet under complete theoretical control. Consequently, we caution that lattice-QCD calculations performed directly at large recoil should not, in general, be extrapolated from unphysical pion masses using a kinematic-independent ansatz for the pion-mass dependence until the essential differences between our results and those of HPChPT have been fully resolved. This point is of immediate phenomenological relevance because some state-of-the-art lattice determinations of the $B\to\pi$ form factors already employ HPChPT, or closely related chiral extrapolation procedures, directly in their analysis. By contrast, the strategy of determining the form factors from lattice calculations at high-$q^2$ where the chiral extrapolation is under better theoretical control, and indeed physical-point calculations are more feasible, by extrapolating them to the full kinematic range using analyticity-based parameterizations, is not negatively affected by our results.

In the future, it might be interesting to extend the construction to the $SU(3)$ case in order to relate the corresponding hadronic functions across the full set of $B \to P$ transitions. It is conceivable that considering several $SU(3)$-related channels simultaneously could provide additional information on the two independent form-factor functions and might help to guide combined chiral extrapolations of lattice data in the low-$q^2$ region. Whether these relations lead to quantitatively useful constraints remains to be established, and we leave a systematic investigation of this possibility to future work.


\section*{Acknowledgements}

We thank Björn O.~Lange for various fruitful discussions on the description of heavy-to-light form factors in the SCET-2 framework.

This work has been supported by the Deutsche Forschungsgemeinschaft (DFG, German Research Foundation) under grant 396021762 - TRR 257, and by Germany’s Excellence Strategy – Cluster of Excellence “Color meets Flavor”, EXC 3107 – Project-ID 533766364.

JJ thanks the Erwin-Schrödinger International Institute for Mathematics and Physics at the University of Vienna for partial support during the Programme `New Paradigms for Harnessing Quantum Field Theory at Colliders', July 27 – August 28, 2026.

\appendix

\section{Redundancy of the current with zero insertions of the axial field}
\label{Appendix:J0_redundancy}


In this section we show that within the family of collinear currents defined in Eq.~\eqref{eq:J2def}, the current with zero insertions of the derivative building block ($a=0$) can be represented in terms of the identity operator and the single-insertion current ($a=1$) as in Eq.~\eqref{eq:J0equiv} below. We begin by subtracting unity to write the zero-insertion current in terms of a total derivative,
\begin{align}
  J^{(1)}_{c,0}(s)-1
  &=
  \Xi_{c,L}(s\bar n)\Xi_{c,L}^\dagger(0)
  -
  \Xi_{c,L}(0)\Xi_{c,L}^\dagger(0)=
  \int_0^s ds_1\,
  \partial_{s_1}
  \left[
    \Xi_{c,L}(s_1\bar n)\Xi_{c,L}^\dagger(0)
  \right] \, . \label{eq:J0equivpre}
\end{align}
Following the definitions in Eq.~\eqref{eq:chiraldef}, it can be shown that the Wilson-line-dressed chiral fields satisfy
\begin{align}
    iD^\mu \Xi_{a,L} &= \Xi_{a,L}(\mathcal{A}_a^\mu +\mathcal{V}_a^\mu), \quad iD^\mu \Xi_{a,R}^\dagger
=  \,\Xi_{a,R}^\dagger (\mathcal{V}_a^\mu-\mathcal{A}_a^\mu) \, .
\end{align}
Since $(\bar{n}\cdot \partial) \, \Xi_{c,L}(s \bar{n})= \partial_s \, \Xi_{c,L}(s \bar{n})$ we can write the derivative in Eq.~\eqref{eq:J0equivpre} as
\begin{align}
  i\partial_s \, \Xi_{c,L}(s\bar n)
  =
   \Xi_{c,L}(s\bar n)\,
  \bar n \cdot \mathcal A_c(s\bar n) - [(\bar n \cdot \ell_c) \  \Xi_{c,L}](s\bar n) \, ,
\end{align}
where we have used $\bar{n}\cdot \mathcal{V}_c=0$. For the matrix elements of interest, we can set $\ell_c^\mu=0$ to simplify our relation to
\begin{align}
  J^{(1)}_{c,0}(s)
  &=
  1-i\int_0^s ds_1 \,
  \Xi_{c,L}(s_1\bar n)\,
  \bar n \cdot \mathcal A_{c}(s_1\bar n)\,
  \Xi_{c,L}^\dagger(0) =1-i\int_0^s ds_1 \, J^{(1)}_{c,1}(s_1, s_1) \, .\label{eq:J0equiv}
\end{align}

Now, since the identity operator does not contribute to a matrix element with one external collinear pion, the zero-insertion current may therefore be dropped as an independent operator in favor of a certain `endpoint configuration' of the one-insertion current, with $s_1=s$, which is subsumed into the general case. This operator redundancy argument is consistent with the result that both currents give the same chiral corrections to this matrix element, which we verified by explicit calculation. Finally, one can show that a similar conclusion holds for expressing the contribution of $J^{(2)}_{c,0}$ to the matrix element of interest in terms of $J^{(2)}_{c,1}$, although the relation takes a slightly different form than Eq.~\eqref{eq:J0equiv}.


\section{Absence of chiral logarithms in the two-point integral}
\label{Appendix:no_chiral_logs}


In Section~\ref{sec:4} we stated that all chiral logarithms arising in $B \to \pi_c$ vertex corrections are associated with tadpole topologies. Here we show that the remaining soft-pion topology and the associated integral $I_2(t)$ defined in Eq.~\eqref{eq:I2} is free of such logarithms, for general $\Delta\geq0$ and $t \neq 0$. The integral can be written as a Georgi-parameter integral
\begin{align}
    I_2(t,m_\pi^2)&=-\frac{\Gamma(\epsilon)\tilde{\mu}^{2\epsilon}}{(1-\epsilon)(4\pi)^{2-\epsilon}} \, t \int_0^\infty d\eta \,e^{it\eta} \big[ A(\eta)+m_\pi^2\big]^{1-\epsilon} \, ,
\end{align}
where $A(\eta)=\eta^2-2\eta \Delta$. Isolating the asymptotics of the integral representation of $I_2(t,m_\pi^2)-I_2(t,0)$ for $\eta \to \infty$, the $\epsilon$-expansion reads
\begin{align}
    I_2(t,m_\pi^2)&=I_2(t,0)-\frac{i m_\pi^2}{(4\pi)^2}\left[ \frac{1}{\epsilon} + \ln(-\mu^2 t^2) \right] + \frac{t}{(4\pi)^2} I_2^{\textrm{rem}}(t,m_\pi^2) + O(\epsilon) \, ,
\end{align}
where the UV-finite remainder is
\begin{align}
    I_2^{\textrm{rem}}(t,m_\pi^2)&= \int_0^\infty d \eta \,e^{it \eta} \left[ A(\eta) \ln \left( 1+ \frac{m_\pi^2}{A(\eta)}\right) + m_\pi^2 \left( \ln \frac{A(\eta) + m_\pi^2}{\eta^2}-1 \right)\right]\, .
\end{align}
To show that $I_2(t)$ is free of chiral logarithms of the form $\Delta^2 \ln m_\pi^2$ and $m_\pi^2 \ln m_\pi^2$, it is sufficient to show that the integral on the RHS of
\begin{align}
    \left.\frac{dI_2^{\textrm{rem}}}{dm_\pi^2}\right|_{m_\pi^2=0}= \int_0^\infty d\eta \,e^{i t \eta} \ln \frac{A(\eta)}{\eta^2} \, ,
\end{align}
is convergent. Indeed, the isolated logarithmic singularities at $\eta=0$ and $\eta=2\Delta$ are integrable, and the oscillatory integrand falls sufficiently rapidly as $\sim e^{it \eta}/\eta$ for $\eta \to \infty$. In the special case $\Delta=0$, one simply has $dI_2^{\textrm{rem}}/dm_\pi^2=0$ at $m_\pi^2=0$.


\bibliographystyle{JHEP}
\bibliography{biblio}

@article{Charles:2015gya,
    author = "Charles, J. and others",
    title = "{Current status of the Standard Model CKM fit and constraints on $\Delta F=2$ New Physics}",
    eprint = "1501.05013",
    archivePrefix = "arXiv",
    primaryClass = "hep-ph",
    reportNumber = "LPT-ORSAY-15-04",
    doi = "10.1103/PhysRevD.91.073007",
    journal = "Phys. Rev. D",
    volume = "91",
    number = "7",
    pages = "073007",
    year = "2015"
}

@article{Bona:2006ah,
    author = "Bona, M. and others",
    collaboration = "UTfit",
    title = "{The Unitarity Triangle Fit in the Standard Model and Hadronic Parameters from Lattice QCD: A Reappraisal after the Measurements of Delta m(s) and BR(B ---{\ensuremath{>}} tau nu(tau))}",
    eprint = "hep-ph/0606167",
    archivePrefix = "arXiv",
    doi = "10.1088/1126-6708/2006/10/081",
    journal = "JHEP",
    volume = "10",
    pages = "081",
    year = "2006"
}

@article{HeavyFlavorAveragingGroupHFLAV:2024ctg,
    author = "Banerjee, Sw. and others",
    collaboration = "Heavy Flavor Averaging Group (HFLAV)",
    title = "{Averages of b-hadron, c-hadron, and {\ensuremath{\tau}}-lepton properties as of 2023}",
    eprint = "2411.18639",
    archivePrefix = "arXiv",
    primaryClass = "hep-ex",
    doi = "10.1103/x87q-tld5",
    journal = "Phys. Rev. D",
    volume = "113",
    number = "1",
    pages = "012008",
    year = "2026"
}

@article{FlavourLatticeAveragingGroupFLAG:2024oxs,
    author = "Aoki, Y. and others",
    collaboration = "Flavour Lattice Averaging Group (FLAG)",
    title = "{FLAG review 2024}",
    eprint = "2411.04268",
    archivePrefix = "arXiv",
    primaryClass = "hep-lat",
    reportNumber = "CERN-TH-2024-192, FERMILAB-PUB-24-0785-T",
    doi = "10.1103/nfzp-p5dn",
    journal = "Phys. Rev. D",
    volume = "113",
    number = "1",
    pages = "014508",
    year = "2026"
}

@article{Dalgic:2006dt,
    author = "Dalgic, Emel and Gray, Alan and Wingate, Matthew and Davies, Christine T. H. and Lepage, G. Peter and Shigemitsu, Junko",
    title = "{B meson semileptonic form-factors from unquenched lattice QCD}",
    eprint = "hep-lat/0601021",
    archivePrefix = "arXiv",
    doi = "10.1103/PhysRevD.75.119906",
    journal = "Phys. Rev. D",
    volume = "73",
    pages = "074502",
    year = "2006",
    note = "[Erratum: Phys.Rev.D 75, 119906 (2007)]"
}

@article{Colquhoun:2015mfa,
    author = "Colquhoun, B. and Dowdall, R. J. and Koponen, J. and Davies, C. T. H. and Lepage, G. P.",
    title = "{B{\textrightarrow}{\ensuremath{\pi}}{\ensuremath{\ell}}{\ensuremath{\nu}} at zero recoil from lattice QCD with physical u/d quarks}",
    eprint = "1510.07446",
    archivePrefix = "arXiv",
    primaryClass = "hep-lat",
    doi = "10.1103/PhysRevD.93.034502",
    journal = "Phys. Rev. D",
    volume = "93",
    number = "3",
    pages = "034502",
    year = "2016"
}

@article{FermilabLattice:2015mwy,
    author = "Bailey, Jon A. and others",
    collaboration = "Fermilab Lattice, MILC",
    title = "{$|V_{ub}|$ from $B\to\pi\ell\nu$ decays and (2+1)-flavor lattice QCD}",
    eprint = "1503.07839",
    archivePrefix = "arXiv",
    primaryClass = "hep-lat",
    reportNumber = "FERMILAB-PUB-15-108-T",
    doi = "10.1103/PhysRevD.92.014024",
    journal = "Phys. Rev. D",
    volume = "92",
    number = "1",
    pages = "014024",
    year = "2015"
}

@article{Flynn:2015mha,
    author = "Flynn, J. M. and Izubuchi, T. and Kawanai, T. and Lehner, C. and Soni, A. and Van de Water, R. S. and Witzel, O.",
    title = "{$B \to \pi \ell \nu$ and $B_s \to K \ell \nu$ form factors and $|V_{ub}|$ from 2+1-flavor lattice QCD with domain-wall light quarks and relativistic heavy quarks}",
    eprint = "1501.05373",
    archivePrefix = "arXiv",
    primaryClass = "hep-lat",
    reportNumber = "FERMILAB-PUB-15-019-T",
    doi = "10.1103/PhysRevD.91.074510",
    journal = "Phys. Rev. D",
    volume = "91",
    number = "7",
    pages = "074510",
    year = "2015"
}

@article{Colquhoun:2022atw,
    author = "Colquhoun, Brian and Hashimoto, Shoji and Kaneko, Takashi and Koponen, Jonna",
    collaboration = "JLQCD",
    title = {{Form factors of B{\textrightarrow}{\ensuremath{\pi}}{\ensuremath{\ell}}{\ensuremath{\nu}} and a determination of |Vub| with M{\"o}bius domain-wall fermions}},
    eprint = "2203.04938",
    archivePrefix = "arXiv",
    primaryClass = "hep-lat",
    reportNumber = "KEK-CP-382",
    doi = "10.1103/PhysRevD.106.054502",
    journal = "Phys. Rev. D",
    volume = "106",
    number = "5",
    pages = "054502",
    year = "2022"
}

@article{Bouchard:2014ypa,
    author = "Bouchard, C. M. and Lepage, G. Peter and Monahan, Christopher and Na, Heechang and Shigemitsu, Junko",
    title = "{$B_s \to K \ell \nu$ form factors from lattice QCD}",
    eprint = "1406.2279",
    archivePrefix = "arXiv",
    primaryClass = "hep-lat",
    doi = "10.1103/PhysRevD.90.054506",
    journal = "Phys. Rev. D",
    volume = "90",
    pages = "054506",
    year = "2014"
}

@article{FermilabLattice:2019ikx,
    author = "Bazavov, Alexei and others",
    collaboration = "Fermilab Lattice, MILC",
    title = "{$B_s\to K\ell\nu$ decay from lattice QCD}",
    eprint = "1901.02561",
    archivePrefix = "arXiv",
    primaryClass = "hep-lat",
    reportNumber = "FERMILAB-PUB-19-005-T",
    doi = "10.1103/PhysRevD.100.034501",
    journal = "Phys. Rev. D",
    volume = "100",
    number = "3",
    pages = "034501",
    year = "2019"
}

@article{Flynn:2023nhi,
    author = {Flynn, J. M. and Hill, R. C. and J{\"u}ttner, A. and Soni, A. and Tsang, J. T. and Witzel, O.},
    collaboration = "RBC/UKQCD",
    title = "{Exclusive semileptonic Bs{\textrightarrow}K{\ensuremath{\ell}}{\ensuremath{\nu}} decays on the lattice}",
    eprint = "2303.11280",
    archivePrefix = "arXiv",
    primaryClass = "hep-lat",
    reportNumber = "CERN-TH-2023-046, FERMILAB-PUB-23-115-V, P3H-23-017, SI-HEP-2023-06",
    doi = "10.1103/PhysRevD.107.114512",
    journal = "Phys. Rev. D",
    volume = "107",
    number = "11",
    pages = "114512",
    year = "2023"
}

@article{Beneke:1999br,
    author = "Beneke, M. and Buchalla, G. and Neubert, M. and Sachrajda, Christopher T.",
    title = "{QCD factorization for B ---{\ensuremath{>}} pi pi decays: Strong phases and CP violation in the heavy quark limit}",
    eprint = "hep-ph/9905312",
    archivePrefix = "arXiv",
    reportNumber = "SLAC-PUB-8146, CERN-TH-99-126, SHEP-99-04",
    doi = "10.1103/PhysRevLett.83.1914",
    journal = "Phys. Rev. Lett.",
    volume = "83",
    pages = "1914--1917",
    year = "1999"
}

@article{Beneke:2000ry,
    author = "Beneke, M. and Buchalla, G. and Neubert, M. and Sachrajda, Christopher T.",
    title = "{QCD factorization for exclusive, nonleptonic B meson decays: General arguments and the case of heavy light final states}",
    eprint = "hep-ph/0006124",
    archivePrefix = "arXiv",
    reportNumber = "CERN-TH-2000-159, CLNS-00-1675, PITHA-00-06, SHEP-00-06",
    doi = "10.1016/S0550-3213(00)00559-9",
    journal = "Nucl. Phys. B",
    volume = "591",
    pages = "313--418",
    year = "2000"
}

@article{Ali:1999mm,
    author = "Ali, Ahmed and Ball, Patricia and Handoko, L. T. and Hiller, G.",
    title = "{A Comparative study of the decays $B \to$ ($K$, $K^{*)} \ell^+ \ell^-$ in standard model and supersymmetric theories}",
    eprint = "hep-ph/9910221",
    archivePrefix = "arXiv",
    reportNumber = "SLAC-PUB-8269, DESY-99-146, CERN-TH-99-298, LNF-99-026-P",
    doi = "10.1103/PhysRevD.61.074024",
    journal = "Phys. Rev. D",
    volume = "61",
    pages = "074024",
    year = "2000"
}

@article{Beneke:2001at,
    author = "Beneke, M. and Feldmann, T. and Seidel, D.",
    title = "{Systematic approach to exclusive $B \to  V l^+ l^-$, $V \gamma$ decays}",
    eprint = "hep-ph/0106067",
    archivePrefix = "arXiv",
    reportNumber = "PITHA-01-05",
    doi = "10.1016/S0550-3213(01)00366-2",
    journal = "Nucl. Phys. B",
    volume = "612",
    pages = "25--58",
    year = "2001"
}

@article{Belle:2010hep,
    author = "Ha, H. and others",
    collaboration = "Belle",
    title = "{Measurement of the decay $B^0\to\pi^-\ell^+\nu$ and determination of $|V_{ub}|$}",
    eprint = "1012.0090",
    archivePrefix = "arXiv",
    primaryClass = "hep-ex",
    reportNumber = "BELLE-PREPRINT-2010-22, KEK-PREPRINT-2010-37",
    doi = "10.1103/PhysRevD.83.071101",
    journal = "Phys. Rev. D",
    volume = "83",
    pages = "071101",
    year = "2011"
}

@article{Belle:2013hlo,
    author = "Sibidanov, A. and others",
    collaboration = "Belle",
    title = "{Study of Exclusive $B \to X_u \ell \nu$ Decays and Extraction of $\|V_{ub}\|$ using Full Reconstruction Tagging at the Belle Experiment}",
    eprint = "1306.2781",
    archivePrefix = "arXiv",
    primaryClass = "hep-ex",
    reportNumber = "BELLE-PREPRINT-2013-9, KEK-PREPRINT-2013-8",
    doi = "10.1103/PhysRevD.88.032005",
    journal = "Phys. Rev. D",
    volume = "88",
    number = "3",
    pages = "032005",
    year = "2013"
}

@article{BaBar:2010efp,
    author = "del Amo Sanchez, P. and others",
    collaboration = "BaBar",
    title = "{Study of $B \to \pi \ell \nu$ and $B \to \rho \ell \nu$ Decays and Determination of $|V_{ub}|$}",
    eprint = "1005.3288",
    archivePrefix = "arXiv",
    primaryClass = "hep-ex",
    reportNumber = "SLAC-PUB-14106, BABAR-PUB-09-037",
    doi = "10.1103/PhysRevD.83.032007",
    journal = "Phys. Rev. D",
    volume = "83",
    pages = "032007",
    year = "2011"
}

@article{BaBar:2012thb,
    author = "Lees, J. P. and others",
    collaboration = "BaBar",
    title = "{Branching fraction and form-factor shape measurements of exclusive charmless semileptonic B decays, and determination of $|V_{ub}|$}",
    eprint = "1208.1253",
    archivePrefix = "arXiv",
    primaryClass = "hep-ex",
    reportNumber = "BABAR-PUB12-015, SLAC-PUB-15208",
    doi = "10.1103/PhysRevD.86.092004",
    journal = "Phys. Rev. D",
    volume = "86",
    pages = "092004",
    year = "2012"
}

@article{LHCb:2020ist,
    author = "Aaij, R. and others",
    collaboration = "LHCb",
    title = "{First observation of the decay $B_s^0 \to K^-\mu^+\nu_\mu$ and Measurement of $|V_{ub}|/|V_{cb}|$}",
    eprint = "2012.05143",
    archivePrefix = "arXiv",
    primaryClass = "hep-ex",
    reportNumber = "LHCb-PAPER-2020-038, CERN-EP-2020-224, CERN-EP-2020-224",
    doi = "10.1103/PhysRevLett.126.081804",
    journal = "Phys. Rev. Lett.",
    volume = "126",
    number = "8",
    pages = "081804",
    year = "2021"
}

@article{Wise:1992hn,
    author = "Wise, Mark B.",
    title = "{Chiral perturbation theory for hadrons containing a heavy quark}",
    reportNumber = "CALT-68-1765",
    doi = "10.1103/PhysRevD.45.R2188",
    journal = "Phys. Rev. D",
    volume = "45",
    number = "7",
    pages = "R2188",
    year = "1992"
}

@article{Yan:1992gz,
    author = "Yan, Tung-Mow and Cheng, Hai-Yang and Cheung, Chi-Yee and Lin, Guey-Lin and Lin, Y. C. and Yu, Hoi-Lai",
    title = "{Heavy quark symmetry and chiral dynamics}",
    reportNumber = "CLNS-92-1138, IP-ASTP-03-92",
    doi = "10.1103/PhysRevD.46.1148",
    journal = "Phys. Rev. D",
    volume = "46",
    pages = "1148--1164",
    year = "1992",
    note = "[Erratum: Phys.Rev.D 55, 5851 (1997)]"
}

@article{Burdman:1992gh,
    author = "Burdman, Gustavo and Donoghue, John F.",
    title = "{Union of chiral and heavy quark symmetries}",
    reportNumber = "UMHEP-365",
    doi = "10.1016/0370-2693(92)90068-F",
    journal = "Phys. Lett. B",
    volume = "280",
    pages = "287--291",
    year = "1992"
}

@article{Boyd:1994pa,
    author = "Boyd, C. Glenn and Grinstein, Benjamin",
    title = "{Chiral and heavy quark symmetry violation in B decays}",
    eprint = "hep-ph/9402340",
    archivePrefix = "arXiv",
    reportNumber = "UCSD-PTH-93-46, SMU-HEP-94-03, SSCL-PREPRINT-532",
    doi = "10.1016/S0550-3213(95)00005-4",
    journal = "Nucl. Phys. B",
    volume = "442",
    pages = "205--227",
    year = "1995"
}

@article{Casalbuoni:1996pg,
    author = "Casalbuoni, R. and Deandrea, A. and Di Bartolomeo, N. and Gatto, Raoul and Feruglio, F. and Nardulli, G.",
    title = "{Phenomenology of heavy meson chiral Lagrangians}",
    eprint = "hep-ph/9605342",
    archivePrefix = "arXiv",
    reportNumber = "UGVA-DPT-1996-05-928, BARI-TH-96-237",
    doi = "10.1016/S0370-1573(96)00027-0",
    journal = "Phys. Rept.",
    volume = "281",
    pages = "145--238",
    year = "1997"
}

@article{Colangelo:2000dp,
    author = "Colangelo, Pietro and Khodjamirian, Alexander",
    editor = "Shifman, M. and Ioffe, Boris",
    title = "{QCD sum rules, a modern perspective}",
    eprint = "hep-ph/0010175",
    archivePrefix = "arXiv",
    reportNumber = "CERN-TH-2000-296, BARI-TH-2000-394",
    doi = "10.1142/9789812810458_0033",
    pages = "1495--1576",
    month = "10",
    year = "2000"
}

@article{Khodjamirian:2023wol,
    author = "Khodjamirian, Alexander and Meli{\'c}, Bla{\v{z}}enka and Wang, Yu-Ming",
    title = "{A guide to the QCD light-cone sum rules for $b$-quark decays}",
    eprint = "2311.08700",
    archivePrefix = "arXiv",
    primaryClass = "hep-ph",
    reportNumber = "SI-HEP-2023-25, P3H-23-087; RBI-ThPhys-2023-37",
    doi = "10.1140/epjs/s11734-023-01046-6",
    journal = "Eur. Phys. J. ST",
    volume = "233",
    number = "2",
    pages = "271--298",
    year = "2024"
}

@article{Beneke:2000wa,
    author = "Beneke, M. and Feldmann, T.",
    title = "{Symmetry breaking corrections to heavy to light B meson form-factors at large recoil}",
    eprint = "hep-ph/0008255",
    archivePrefix = "arXiv",
    reportNumber = "PITHA-00-20",
    doi = "10.1016/S0550-3213(00)00585-X",
    journal = "Nucl. Phys. B",
    volume = "592",
    pages = "3--34",
    year = "2001"
}

@article{Bauer:2000yr,
    author = "Bauer, Christian W. and Fleming, Sean and Pirjol, Dan and Stewart, Iain W.",
    title = "{An Effective field theory for collinear and soft gluons: Heavy to light decays}",
    eprint = "hep-ph/0011336",
    archivePrefix = "arXiv",
    reportNumber = "UCSD-PTH-00-28",
    doi = "10.1103/PhysRevD.63.114020",
    journal = "Phys. Rev. D",
    volume = "63",
    pages = "114020",
    year = "2001"
}

@article{Bauer:2001yt,
    author = "Bauer, Christian W. and Pirjol, Dan and Stewart, Iain W.",
    title = "{Soft collinear factorization in effective field theory}",
    eprint = "hep-ph/0109045",
    archivePrefix = "arXiv",
    reportNumber = "UCSD-PTH-01-15",
    doi = "10.1103/PhysRevD.65.054022",
    journal = "Phys. Rev. D",
    volume = "65",
    pages = "054022",
    year = "2002"
}

@article{Beneke:2002ph,
    author = "Beneke, M. and Chapovsky, A. P. and Diehl, M. and Feldmann, T.",
    title = "{Soft collinear effective theory and heavy to light currents beyond leading power}",
    eprint = "hep-ph/0206152",
    archivePrefix = "arXiv",
    reportNumber = "PITHA-02-09",
    doi = "10.1016/S0550-3213(02)00687-9",
    journal = "Nucl. Phys. B",
    volume = "643",
    pages = "431--476",
    year = "2002"
}

@article{Beneke:2003pa,
    author = "Beneke, M. and Feldmann, T.",
    title = "{Factorization of heavy to light form-factors in soft collinear effective theory}",
    eprint = "hep-ph/0311335",
    archivePrefix = "arXiv",
    reportNumber = "PITHA-03-11, CERN-TH-2003-286",
    doi = "10.1016/j.nuclphysb.2004.02.033",
    journal = "Nucl. Phys. B",
    volume = "685",
    pages = "249--296",
    year = "2004"
}

@article{Lange:2003pk,
    author = "Lange, Bjorn O. and Neubert, Matthias",
    title = "{Factorization and the soft overlap contribution to heavy to light form-factors}",
    eprint = "hep-ph/0311345",
    archivePrefix = "arXiv",
    reportNumber = "CLNS-03-1849",
    doi = "10.1016/j.nuclphysb.2005.06.019",
    journal = "Nucl. Phys. B",
    volume = "690",
    pages = "249--278",
    year = "2004",
    note = "[Erratum: Nucl.Phys.B 723, 201--202 (2005)]"
}

@article{Hill:2002vw,
    author = "Hill, Richard J. and Neubert, Matthias",
    title = "{Spectator interactions in soft collinear effective theory}",
    eprint = "hep-ph/0211018",
    archivePrefix = "arXiv",
    reportNumber = "SLAC-PUB-9549, CLNS-02-1803",
    doi = "10.1016/S0550-3213(03)00116-0",
    journal = "Nucl. Phys. B",
    volume = "657",
    pages = "229--256",
    year = "2003"
}

@article{Chen:2003fp,
    author = "Chen, Jiunn-Wei and Stewart, Iain W.",
    title = "{Model independent results for SU(3) violation in light cone distribution functions}",
    eprint = "hep-ph/0311285",
    archivePrefix = "arXiv",
    reportNumber = "MIT-CTP-3445",
    doi = "10.1103/PhysRevLett.92.202001",
    journal = "Phys. Rev. Lett.",
    volume = "92",
    pages = "202001",
    year = "2004"
}

@article{Flynn:2008tg,
    author = "Flynn, J. M. and Sachrajda, C. T.",
    collaboration = "RBC, UKQCD",
    title = "{SU(2) chiral perturbation theory for K(l3) decay amplitudes}",
    eprint = "0809.1229",
    archivePrefix = "arXiv",
    primaryClass = "hep-ph",
    reportNumber = "SHEP-08-26",
    doi = "10.1016/j.nuclphysb.2008.12.001",
    journal = "Nucl. Phys. B",
    volume = "812",
    pages = "64--80",
    year = "2009"
}

@article{Bijnens:2009yr,
    author = "Bijnens, Johan and Celis, Alejandro",
    title = "{K ---{\ensuremath{>}} pi pi Decays in SU(2) Chiral Perturbation Theory}",
    eprint = "0906.0302",
    archivePrefix = "arXiv",
    primaryClass = "hep-ph",
    reportNumber = "LU-TP-09-14",
    doi = "10.1016/j.physletb.2009.09.048",
    journal = "Phys. Lett. B",
    volume = "680",
    pages = "466--470",
    year = "2009"
}

@article{Bijnens:2010ws,
    author = "Bijnens, Johan and Jemos, Ilaria",
    title = "{Hard Pion Chiral Perturbation Theory for $B\to\pi$ and $D\to\pi$ Formfactors}",
    eprint = "1006.1197",
    archivePrefix = "arXiv",
    primaryClass = "hep-ph",
    reportNumber = "LU-TP-10-16",
    doi = "10.1016/j.nuclphysb.2010.06.021",
    journal = "Nucl. Phys. B",
    volume = "840",
    pages = "54--66",
    year = "2010",
    note = "[Erratum: Nucl.Phys.B 844, 182--183 (2011)]"
}

@article{Bijnens:2010jg,
    author = "Bijnens, Johan and Jemos, Ilaria",
    title = "{Vector Formfactors in Hard Pion Chiral Perturbation Theory}",
    eprint = "1011.6531",
    archivePrefix = "arXiv",
    primaryClass = "hep-ph",
    reportNumber = "LU-TP-10-27",
    doi = "10.1016/j.nuclphysb.2010.12.012",
    journal = "Nucl. Phys. B",
    volume = "846",
    pages = "145--166",
    year = "2011"
}

@article{Colangelo:2012ew,
    author = "Colangelo, G. and Procura, M. and Rothen, L. and Stucki, R. and Tarrus Castella, J.",
    title = "{On the factorization of chiral logarithms in the pion form factors}",
    eprint = "1208.0498",
    archivePrefix = "arXiv",
    primaryClass = "hep-ph",
    doi = "10.1007/JHEP09(2012)081",
    journal = "JHEP",
    volume = "09",
    pages = "081",
    year = "2012"
}

@article{Grinstein:2005ry,
    author = "Grinstein, Benjamin and Pirjol, Dan",
    title = "{Chiral symmetry and exclusive B decays in the SCET}",
    eprint = "hep-ph/0501237",
    archivePrefix = "arXiv",
    reportNumber = "UCSD-PTH-05-1, MIT-CTP-3592",
    doi = "10.1016/j.physletb.2005.04.024",
    journal = "Phys. Lett. B",
    volume = "615",
    pages = "213--220",
    year = "2005"
}

@article{Callan:1969sn,
    author = "Callan, Jr., Curtis G. and Coleman, Sidney R. and Wess, J. and Zumino, Bruno",
    title = "{Structure of phenomenological Lagrangians. 2.}",
    doi = "10.1103/PhysRev.177.2247",
    journal = "Phys. Rev.",
    volume = "177",
    pages = "2247--2250",
    year = "1969"
}

@article{Bando:1987br,
    author = "Bando, Masako and Kugo, Taichiro and Yamawaki, Koichi",
    title = "{Nonlinear Realization and Hidden Local Symmetries}",
    reportNumber = "DPNU-87-63, AICHI-1, KUNS-903",
    doi = "10.1016/0370-1573(88)90019-1",
    journal = "Phys. Rept.",
    volume = "164",
    pages = "217--314",
    year = "1988"
}

@article{Coleman:1969sm,
    author = "Coleman, Sidney R. and Wess, J. and Zumino, Bruno",
    title = "{Structure of phenomenological Lagrangians. 1.}",
    doi = "10.1103/PhysRev.177.2239",
    journal = "Phys. Rev.",
    volume = "177",
    pages = "2239--2247",
    year = "1969"
}

@article{Honerkamp:1971sh,
    author = "Honerkamp, J.",
    title = "{Chiral multiloops}",
    doi = "10.1016/0550-3213(72)90299-4",
    journal = "Nucl. Phys. B",
    volume = "36",
    pages = "130--140",
    year = "1972"
}

@article{Meetz:1969as,
    author = "Meetz, K.",
    title = "{Realization of chiral symmetry in a curved isospin space}",
    doi = "10.1063/1.1664881",
    journal = "J. Math. Phys.",
    volume = "10",
    pages = "589--593",
    year = "1969"
}

@article{Gasser:1987rb,
    author = "Gasser, J. and Sainio, M. E. and Svarc, A.",
    title = "{Nucleons with chiral loops}",
    reportNumber = "BUTP-87-17",
    doi = "10.1016/0550-3213(88)90108-3",
    journal = "Nucl. Phys. B",
    volume = "307",
    pages = "779--853",
    year = "1988"
}

@article{Fleischer:1992tn,
    author = "Fleischer, Robert",
    title = "{Chiral logarithmic corrections to heavy to light semileptonic decays of the type H ---{\ensuremath{>}} pi e neutrino}",
    reportNumber = "TUM-T31-34-92",
    doi = "10.1016/0370-2693(93)90059-Q",
    journal = "Phys. Lett. B",
    volume = "303",
    pages = "147--151",
    year = "1993"
}

@article{Falk:1993fr,
    author = "Falk, Adam F. and Grinstein, Benjamin",
    title = "{Anti-B ---{\ensuremath{>}} Anti-K e+ e- in Chiral Perturbation Theory}",
    eprint = "hep-ph/9306310",
    archivePrefix = "arXiv",
    reportNumber = "SLAC-PUB-6237, SSCL-PREPRINT-484",
    doi = "10.1016/0550-3213(94)90554-1",
    journal = "Nucl. Phys. B",
    volume = "416",
    pages = "771--785",
    year = "1994"
}

@article{Becher:2003qh,
    author = "Becher, Thomas and Hill, Richard J. and Neubert, Matthias",
    title = "{Soft collinear messengers: A New mode in soft collinear effective theory}",
    eprint = "hep-ph/0308122",
    archivePrefix = "arXiv",
    reportNumber = "SLAC-PUB-10098, CLNS-03-1836",
    doi = "10.1103/PhysRevD.69.054017",
    journal = "Phys. Rev. D",
    volume = "69",
    pages = "054017",
    year = "2004"
}

@article{Gasser:1984gg,
    author = "Gasser, J. and Leutwyler, H.",
    title = "{Chiral Perturbation Theory: Expansions in the Mass of the Strange Quark}",
    reportNumber = "CERN-TH-3798",
    doi = "10.1016/0550-3213(85)90492-4",
    journal = "Nucl. Phys. B",
    volume = "250",
    pages = "465--516",
    year = "1985"
}

@article{Chen:2005js,
    author = "Chen, Jiunn-Wei and Tsai, Hung-Ming and Weng, Ke-Chuan",
    title = "{Model-independent results for SU(3) violation in twist-3 light-cone distribution functions}",
    eprint = "hep-ph/0511036",
    archivePrefix = "arXiv",
    doi = "10.1103/PhysRevD.73.054010",
    journal = "Phys. Rev. D",
    volume = "73",
    pages = "054010",
    year = "2006"
}

@article{Goity:1992tp,
    author = "Goity, J. L.",
    title = "{Chiral perturbation theory for SU(3) breaking in heavy meson systems}",
    eprint = "hep-ph/9206230",
    archivePrefix = "arXiv",
    reportNumber = "CEBAF-TH-92-16",
    doi = "10.1103/PhysRevD.46.3929",
    journal = "Phys. Rev. D",
    volume = "46",
    pages = "3929--3936",
    year = "1992"
}

@article{Grinstein:1992qt,
    author = "Grinstein, Benjamin and Jenkins, Elizabeth Ellen and Manohar, Aneesh V. and Savage, Martin J. and Wise, Mark B.",
    title = "{Chiral perturbation theory for f D(s) / f D and B B(s) / B B}",
    eprint = "hep-ph/9204207",
    archivePrefix = "arXiv",
    reportNumber = "UCSD-PTH-92-05, CALT-68-1768, SSCL-PREPRINT-025",
    doi = "10.1016/0550-3213(92)90248-A",
    journal = "Nucl. Phys. B",
    volume = "380",
    pages = "369--376",
    year = "1992"
}

@article{Ball:2006wn,
    author = "Ball, Patricia and Braun, V. M. and Lenz, A.",
    title = "{Higher-twist distribution amplitudes of the K meson in QCD}",
    eprint = "hep-ph/0603063",
    archivePrefix = "arXiv",
    reportNumber = "IPPP-06-01, DCPT-06-02, CERN-PH-TH-2006-026",
    doi = "10.1088/1126-6708/2006/05/004",
    journal = "JHEP",
    volume = "05",
    pages = "004",
    year = "2006"
}

@article{Braun:2006dg,
    author = "Braun, V. M. and others",
    title = "{Moments of pseudoscalar meson distribution amplitudes from the lattice}",
    eprint = "hep-lat/0606012",
    archivePrefix = "arXiv",
    reportNumber = "DESY-06-091, EDINBURGH-2006-11, LU-ITP-2006-008",
    doi = "10.1103/PhysRevD.74.074501",
    journal = "Phys. Rev. D",
    volume = "74",
    pages = "074501",
    year = "2006"
}

@article{BurgosMarcos:2025xja,
    author = "Burgos Marcos, M. and Reboud, M. and Vos, K. K.",
    title = "{Detailed SU(3) flavour symmetry analysis of charmless two-body B-meson decays including factorizable corrections}",
    eprint = "2504.05209",
    archivePrefix = "arXiv",
    primaryClass = "hep-ph",
    reportNumber = "EOS-2025-01, Nikhef 2025-004",
    doi = "10.1007/JHEP03(2026)227",
    journal = "JHEP",
    volume = "03",
    pages = "227",
    year = "2026"
}

@article{Fang:2026fhl,
    author = "Fang, Wen-Sheng and Huber, Tobias and Li, Xin-Qiang and Malami, Eleftheria and Tetlalmatzi-Xolocotzi, Gilberto",
    title = "{QCD-factorization amplitudes from flavour symmetries: beyond the $SU(3)$ symmetric case}",
    eprint = "2604.19612",
    archivePrefix = "arXiv",
    primaryClass = "hep-ph",
    reportNumber = "SI-HEP-2026-09, P3H-26-029",
    month = "4",
    year = "2026"
}

@article{Becirevic:2003ad,
    author = "Becirevic, Damir and Prelovsek, Sasa and Zupan, Jure",
    title = "{B ---{\ensuremath{>}} pi and B --{\ensuremath{>}} K transitions in partially quenched chiral perturbation theory}",
    eprint = "hep-lat/0305001",
    archivePrefix = "arXiv",
    reportNumber = "LPT-ORSAY-03-19, IJS-TP-6-03",
    doi = "10.1103/PhysRevD.68.074003",
    journal = "Phys. Rev. D",
    volume = "68",
    pages = "074003",
    year = "2003"
}


\end{document}